\documentstyle[12pt]{article}
\def\Al{{\cal A}}
\def\End{{\cal B}}
\def\Di{{\cal D}}
\def\Ideal{{\cal I}}
\def\Fu{{\cal C}}
\def\Hi{{\cal H}}
\def\Mod{{\cal M}}
\def\Pe{{\cal P}}
\def\Lie{{\cal L}}
\def\Ko{{\cal K}}
\def\Clif{{\cal C}\ell}
\def\Kcyc{{\cal T}}
\def\Re{{\bf R}}
\def\Co{{\bf C}}
\def\clif{{\bf c}}
\def\X{{\bf X}}
\def\Y{{\bf Y}}
\def\Ent{{\bf Z}}
\def\Proj{{\bf P}}
\def\dif{{\bf d}}
\def\Id{{\bf 1}}
\def\im{{\bf i}}
\def\Dirac{{\bf D}}
\def\L{{\bf L}}
\def\K{{\bf K}}
\def\J{{\bf J}}
\def\U{{\bf U}}
\def\bu{{\bf u}}
\def\BPsi{{\bf \Psi}}
\def\BPhi{{\bf \Phi}}
\def\chiral{{\bf \Gamma}}
\def\fatl{(\!(}
\def\fatr{)\!)}
\def\BL{\Bigl(\!\Bigl(}
\def\BR{\Bigr)\!\Bigr)}

\def\BRR{\rangle\!\rangle}
\def\ev{\vec e}
\def\part{\vec\partial}
\def\edth{\check\delta\!\!\!/}
\def\be{\begin{equation}}
\def\ee{\end{equation}}
\def\bea{\begin{eqnarray}}
\def\eea{\end{eqnarray}}
\def\beann{\begin{eqnarray*}}
\def\eeann{\end{eqnarray*}}
\begin{document}
%
%%% Titelbladzijde 
\hfill {\bf UFRJ-IF-NCG-1999/2}
%
%\begin{center}{\large\bf This is only a draft!}\end{center}
%
\vskip 1cm
\begin{center} {\Large \bf Connes-Lott model building on the two-sphere}
\vskip 0,5cm {\bf \large J.A.Mignaco${}^{(*)}$
\footnote {partially supported by CNPq,Brasil}, C.Sigaud${}^{(*)}$, A.R.da Silva${}^{(**)}$ and
F.J.Vanhecke${}^{(*)}$\\ {\small E-mail: mignaco@if.ufrj.br, sigaud@if.ufrj.br, beto@dmm.im.ufrj.br
and
 vanhecke@if.ufrj.br }}
\vskip 0.5cm
${}^{(*)}${\it  Instituto de F\'\i sica}, ${}^{(**)}${\it Instituto de Matem\'atica,}\\ {\it UFRJ,
Ilha do Fund\~ao, Rio de Janeiro, Brasil.}\\
\end{center}
\begin{abstract} 
In this work we examine generalized Connes-Lott models, with $\Co\oplus\Co$ as finite algebra, over the two-sphere. 
The Hilbert space of the continuum spectral triple is taken as the space of sections of a twisted
spinor bundle, allowing for nontrivial topological structure (magnetic monopoles). 
The finitely generated projective module over the full algebra is also taken as topologically
non-trivial, which is possible over $S^2$. We also construct a real spectral triple enlarging this
Hilbert space to include "particle" and "anti-particle" fields. 
\end{abstract}
%%%
\vskip 4cm
\noindent PACS numbers : 11.15.-q, 02.40.-k ; \\ Keywords : Noncommutative geometry, Connes-Lott
model, Real spectral triples
%
%%% einde van de titelbladzijde
%
\newpage
\section{Introduction}
\setcounter{equation}{0} In previous work \cite{Migetal1}, we studied the Connes-Lott program with
the complex algebra $\Al_1=\Fu(S^2;\Co)$ of continuous complex-valued functions on the sphere. 
The Hilbert space $\Hi_I$, on which this algebra was represented, consisted of one of the minimal
left ideals $\Ideal_{\pm}$ of the algebra of sections of the Clifford bundle over $S^2$ with a
standard scalar product. On this Hilbert space the Dirac operator was taken as
$\Di_I=\im(\dif-\delta)$ restricted to each of the ideals $\Ideal_\pm$. The projective modules
$\Mod$ over $\Al_1$ were constructed using the Bott projector 
${\bf P}={1\over 2}({\bf 1}+\vec n\cdot\vec\sigma)$, acting on the free module
$\Al_1\oplus\Al_1$.  These modules are classified by the homotopy classes of the mappings $\vec
n:S^2\rightarrow S^2$ i.e. by $\pi_2(S^2)=\Ent$. In Dirac's interpretation, each integer corresponds
to a magnetic monopole at the center of the sphere with magnetic charge $g$ quantised by
\(eg/4\pi=(n/2)\hbar\). 

In the present paper we extend the above analysis of topologically non trivial aspects in  
noncommutative geometry, to the product algebra $\Al=\Al_1\otimes\Al_2$, where $\Al_2=\Co\oplus\Co$. It is
clear that $\Al\simeq\Fu(S^2\times\{a,b\};\Co)$, where $\{a,b\}$ denotes a two-point space, as in
the original Connes-Lott paper \cite{C-L}. \\ In section {\bf \ref{Pensov}} we construct the
Hilbert space on which $\Al_1$ is represented. This Hilbert space $\Hi_{(s)}$  generalizes $\Hi_I$
above, and is made of sections of what we call a Pensov spinor bundle of (integer or semi-integer)
weight $s$, using a taxonomy introduced by Staruszkiewicz \cite{Star}. A generalized Dirac operator
$\Di_{(s)}$ acting on these Pensov spinors is defined and, for $s=\pm1/2$ we recover the K\"{a}hler
spinors of \cite{Migetal1}, while for $s=0$ the usual Dirac spinors are obtained. These spinors may
actually be identified with sections of twisted spinor bundles or, from a more physical viewpoint, as 
usual Dirac spinors interacting with a magnetic monopole of charge $g$ given by \(eg/4\pi\hbar=s\).\\ 
The projective modules over $\Al$, following Connes-Lott, are constructed in
section {\bf\ref{Module}} as $\Mod=\Proj(\Al\oplus\Al)$ where ${\bf P}=\Bigl({\bf P}_a={\bf 1},{\bf
P}_b={1\over 2}({\bf 1}+\vec n_b\cdot\vec\sigma)\Bigr)$. In Connes' work \cite{Con1}, the smooth
manifold is four-dimensional so that, taking the four-sphere $S^4$ as an example, we get mappings $\vec n :
S^4\rightarrow S^2$ classified by $\pi_4(S^2)=\Ent_2$.  However, the local unitary
transformations acting as ${\bf P}_b\rightarrow U^\dagger{\bf P}_b U$ are also classified by homotopy classes
$\pi_4(U(2))=\pi_4(SU(2))=\pi_4(S^3)=\Ent_2$. It follows that\footnote{We are indebted to
prof.Balachandran of Syracuse University for discussions on this point.} all Bott projectors define
modules isomorphic to the module obtained from ${\bf P}_b={1\over 2}({\bf 1}+\sigma_3)$, considered
by Connes. For the two-sphere $S^2$ this does not happen since $\pi_2(S^2)=\Ent$ and
$\pi_2(U(2))=\pi_2(SU(2))=\pi_2(S^3)=\{\Id\}$. In section {\bf\ref{Triple}} we construct the full
spectral triple obtained as the product of the Dirac-Pensov triple for the algebra $\Al_1$ with a
discrete spectral triple for the algebra $\Al_2$. Following again Connes' prescription \`{a} la
lettre, we take the discrete Hilbert space as $\Hi_{dis}=\Co^{N_a}\oplus\Co^{N_b}$ with chirality
$\chi_{dis}$ given as +1 on the a-sector and -1 on the b-sector. The discrete Dirac operator
$\Di_{dis}$ is then the most general hermitian matrix, odd with respect to the grading defined by
$\chi_{dis}$. Eliminating the "junk" in the induced representation of the universal differential
enveloppe $\Omega^{\bullet}(\Al)$, yields bounded operators $\Omega^{\bullet}_D(\Al)$ in
$\Hi=\Hi_{(s)}\otimes\Hi_{dis}$. The standard use of the Dixmier trace and of Connes'trace theorem
allows then to define a scalar product of operators in $\Omega^{\bullet}_D(\Al)$. This scalar
product is used in section {\bf\ref{Yang-Mills-Higgs}} to construct the Yang-Mills-Higgs action.
The main new features in this action, as compared with Connes' result, are the
appearance of an additional monopole potential of strength \(eg/4\pi=(n/2)\hbar\), where $n$ is the
integer characterizing the homotopy class of $\Proj_b$, and the fact that the Higgs doublet is not
globally defined on $S^2$ but transforms as a Pensov field of weight $\pm n/2$. The particle sector 
is examined in section {\bf\ref{Matter}} and a covariant Dirac operator $\Di_\nabla$ acting on 
$\Hi_p=\Mod\otimes_\Al\Hi$ is defined. Here, the novelty is that, whilst the "a-doublet" continues 
as a doublet of Pensov spinors of weight $s$, the "b-singlet" metamorphoses in a Pensov spinor of 
weight $s+n/2$. If one should insist on a comparison with thestandard electroweak model on $S^2$, 
this would mean that right-handed electrons see a different magnetic monopole than the left-handed 
and this is not really welcome. In section {\bf\ref{Real Triples}} we introduce a real Dirac-Pensov 
spectral triple by doubling the Hilbert space as $\Hi_{1}=\Hi_{(s)}\oplus\Hi_{(-s)}$. It is seen that, 
with the same $\Hi_{dis}$ as before, it is not possible to define a  real structure. However, a more 
general discrete Hilbert space $\Hi_{2}=\Co^{N_{aa}}\oplus\Co^{N_{ab}}\oplus\Co^{N_{ba}}\oplus\Co^{N_{bb}}$, 
as considered in\cite{Kraj,Pasch}, allows for  the construction of a real structure on
$\Hi_{new}=\Hi_{1}\otimes\Hi_{2}$. The covariant Dirac operator on 
$\Mod\otimes_\Al\Hi_{new}\otimes_{\Al}\Mod^*$ can also be defined and it is furthermore seen that, 
with the use of such a non trivial projective module, the abelian gauge fields are not slain, 
as they are when $\Mod=\Al$\cite{Var}. 

Clearly this model building led us far from a toy electroweak model. The main purpose however 
is not to reproduce such a model on the two-sphere, but rather to examine some of the
topologically nontrivial structures in model building with the simplest manifold allowing for such
possibilities.
\newpage
\section{The Hilbert space of Pensov spinors on $S^2$}
\label{Pensov}\setcounter{equation}{0} 
The standard atlas of the two-sphere $S^2=\{(x,y,z)\in \Re^3 \mid x^2+y^2+z^2=1\}$ consists of two 
charts, the boreal, $H_B=\{(x,y,z)\in S^2\mid -1<z\leq +1\}$, and austral chart, 
$H_A=\{(x,y,z)\in S^2\mid -1\leq z < +1\}$, with coordinates :
\beann
\zeta_B&=&\xi^1_B+\im \xi^2_B=+ {x+\im y \over 1+z}\;\hbox{ in $H_B$}\;,\\
\zeta_A&=&\xi^1_A+\im \xi^2_A=- {x-\im y \over 1-z}\;\hbox{ in $H_A$}\;.
\eeann 
In the overlap $H_B\cap H_A$, they are related by $\zeta_A\,\zeta_B=-1$  and the usual spherical
coordinates $(\theta,\varphi)$, given by $\zeta_B=-1/\zeta_A=\tan \theta/2\;\exp \im\varphi$, are 
nonsingular. In each chart, dual coordinate bases of the complexified tangent and cotangent spaces
are :
\beann
&&\biggl\{\part={\partial\over\partial\zeta}= {1\over 2}\biggl({\partial\over\partial\xi^1}
-\im{\partial\over\partial\xi^2}\biggr) ,\part^{\;*}={\partial\over\partial\zeta^*}= 
{1\over 2}\biggl({\partial\over\partial\xi^1}+\im {\partial\over\partial\xi^2}\biggr)\biggr\}\\
&&
\biggl\{d\zeta=d\xi^1+\im d\xi^2,d\zeta^*=d\xi^1-\im d\xi^2\biggr\}\;.\eeann 
In $H_B\cap H_A$ they are related by 
\beann
\left(
\begin{array}{cc} 
\part_A & \part_A^{\;*} \end{array}
\right)
& = & \left(\begin{array}{cc}
\part_B & \part_B^{\;*}\end{array}\right)\;
\left(
\begin{array}{cc}
{\zeta_B}^2 & 0 \\
0 & {\zeta_B^*}^2\end{array}\right)\;,\\
\left(\begin{array}{c}
d\zeta_B \\ d\zeta_B^*\end{array}\right)
& = &
\left(\begin{array}{cc}
{\zeta_B}^2 & 0 \\
0 & {\zeta_B^*}^2\end{array}\right)
\left(\begin{array}{c}
d\zeta_A \\ d\zeta_A^*\end{array}\right)\;.\eeann
The euclidean metric in $\Re^3$ induces a metric on the sphere :
\[{\bf g}={4\over q^2}\delta_{ij}\,d\xi^i\otimes d\xi^j=
{2\over q^2}\biggl(d\zeta^*\otimes d\zeta + d\zeta\otimes d\zeta^*\biggr)\;,
\]
where $q= 1+\vert\zeta\vert^2$.
Real and complex Zweibein fields are given by:
\be\label{Hil1}
\biggl\{\theta^i={2\over q}d\xi^i\;;i=1,2\biggr\}\;\hbox{and}\;
\biggl\{\theta={2\over q}d\zeta,\theta^*={2\over q}d\zeta^*\biggr\}\;,\ee
with duals
\[\biggr\{\ev_i={q\over 2}{\partial\over\partial\xi^i}\;;i=1,2\biggr\}
\;\hbox{and}\;
\biggl\{\ev={q\over 2}{\partial\over\partial\zeta}\;,\;
\ev^*={q\over 2}{\partial\over\partial\zeta^*}\biggr\}\;.\]
A rotation of the real Zweibein by an angle $\alpha$ :
\[\left(\begin{array}{c}
\theta^1\\ \theta^2\end{array}\right)\Rightarrow
\left(\begin{array}{c}
\tilde\theta^1\\ \tilde\theta^2\end{array}\right)=
\left(\begin{array}{cc}
\cos\alpha & \sin\alpha \\ -\sin\alpha & \cos\alpha\end{array}\right)\;
\left(\begin{array}{c}
\theta^1 \\
\theta^2\end{array}\right)\;,\]  
becomes diagonal for the complex Zweibein :
\be\label{Hil2}
\left(\begin{array}{c}
\theta\\ \theta^*\end{array}\right)\Rightarrow
\left(\begin{array}{c}
\tilde\theta\\ \tilde\theta^*\end{array}\right)=
\left(\begin{array}{cc}
\exp(-\im\alpha) & 0 \\ 0 & \exp(\im\alpha)\end{array}\right)\;
\left(\begin{array}{c}
\theta \\ \theta^*\end{array}\right)\;.\ee 
This means that the complexified cotangent bundle $(T^*S^2)^\Co$ splits, in an $SO(2)$ invariant way,
into the direct sum of two line bundles $(T^*S^2)^{\prime}$ and $(T^*S^2)^{\prime\prime}$ with 
one-dimensional local bases of sections given by $\{\theta\}$ and $\{\theta^*\}$.\\
In the overlap $H_B\cap H_A$, the Zweibein in $H_A$ and in $H_B$ are related by :
\be\label{Hil3}
\theta_A=(c_{AB})^{-1}\;\theta_B\;,\;\theta_A^*=c_{AB}\;\theta_B^*\;,\ee
whith the transition function \(c_{AB}=\zeta_B/\zeta_B^*=\zeta_A^*/\zeta_A =\exp(2\im\varphi)\),
$\varphi$ being the azimuthal angle, well defined (modulo $2\pi$) in $H_B\cap H_A$.\\ 
Sections of $(T^*S^2)^{\prime}$ and $(T^*S^2)^{\prime\prime}$ are written as
\(\Sigma^{\prime}= \sigma^{(+1)}\;\theta\) and 
\(\Sigma^{\prime\prime}=\sigma^{(-1)}\;\theta^*\) such that, in $H_B\cap H_A$, 
\[{\sigma^{(\pm 1)}}_{\mid A}=(c_{AB})^{\pm 1}{\sigma^{(\pm 1)}}_{\mid B}\;.\]
Following Staruszkiewicz \cite{Star}, who refers to Pensov, we call such a field a Pensov
scalar of weight $(\pm 1)$. The question is now adressed to define Pensov scalars of 
weight $s$ on $S^2$. In general this would require a cocycle condition on transition functions in
triple overlaps. However, since the sphere is covered by only two charts, it is enough that the
overlap equation 
\[{\sigma^{(s)}}_{\mid A}=(c_{AB})^s {\sigma^{(s)}}_{\mid B}\]
be well defined. Now, $(c_{AB})^s=\exp(2\im s\varphi)$ is well defined when $2s$ takes integer 
values\footnote{The integer $2s$ can be identified with the integer representing an element of the
second \v{C}ech cohomology group of $S^2$ with integer values, $\check{H}^2(S^2,\Ent)=\Ent$,
classifying the line bundles over the sphere $S^2$.}.\\  The corresponding line bundle\footnote{In
the sequel, abusing the notation, we shall denote bundles and their spaces of sections by the same
symbol.} will be denoted by $\Pe^{(s)}$.\\ A local basis of its sections in $H_B$ is denoted by
${\Theta^{(s)}}_{\mid B}$, and in $H_A$ by ${\Theta^{(s)}}_{\mid A}$.  They are related in $H_B\cap
H_A$ by the generalisation of (\ref{Hil3}):
\[{\Theta^{(s)}}_{\mid A}=(c_{AB})^{-s}{\Theta^{(s)}}_{\mid B}\;\]
and a local section is given by :
\(\Sigma^{(s)}={\sigma^{(s)}}\;{\Theta^{(s)}}\;\). \\
On $S^2$ with metric ${\bf g}=\delta_{ij}\theta^i\otimes\theta^j$, the Levi-Civita connection reads
\[\nabla^{LC} \theta^i=-{\bf \Gamma}^i_{\;j}\otimes\theta^j
=-{\Gamma_k^{\;\;i}}_j\;\theta^k\otimes\theta^j\;,\]
where
\[{\Gamma_k^{\;\;i}}_j =
-{1\over 2}\{{\delta^i}_k {\partial q\over\partial\xi^j}-
\delta^{i\ell}{\partial q\over\partial\xi^\ell}\delta_{kj}\}\;.\] 
In terms of the complexified Zweibein (\ref{Hil1}), we write :
\be\label{Hil4}
\nabla^{LC} \theta=-{\bf \Gamma}\otimes\theta\;,\;
\nabla^{LC} \theta^*=-{\bf \Gamma}^*\otimes\theta^*\;,\ee
where 
\[{\bf\Gamma}=-{\bf\Gamma}^*=-{1\over 2}
\{{\partial q\over\partial\zeta}\theta\;-{\partial q\over\partial\zeta^*}\;\theta^*\}=
-{1\over 2}\{\zeta^*\theta-\zeta\theta^*\}\;.\]
It is easy to see that \(\nabla^{LC}\Theta^{(s)}=-s{\bf \Gamma}\otimes\Theta^{(s)}\)
defines a connection in the module of Pensov $s$-scalars generalising (\ref{Hil4}) above. 
This connection maps $\Pe^{(s)}$ in $(T^*(S^2))^\Co\otimes\Pe^{(s)}$. Now the
space of complex-valued one-forms $(T^*(S^2))^\Co$ is isomorphic to $\Pe^{(+1)}\oplus\Pe^{(-1)}$,
so that $\nabla^{LC}$ is actually a mapping :
\[\nabla^{LC}:\Pe^{(s)}\mapsto\Pe^{(s+1)}\oplus\Pe^{(s-1)}:
\Sigma^{(s)}\mapsto
\nabla^{LC} \Sigma^{(s)}=
\biggl(\dif\sigma^{(s)}-s{\bf \Gamma}\sigma^{(s)}\biggr)\otimes\Theta^{(s)}\;.\]
Projecting $\nabla^{LC}\Psi^{(s)}$ on each term in the sum $\Pe^{(s+1)}\oplus\Pe^{(s-1)}$ we obtain
:
\[\nabla^{LC}\Psi^{(s)} ={1\over 2}\biggl(\edth_s\sigma^{(s)}\Theta^{(s+1)}+
\edth_s^\dagger\sigma^{(s)}\Theta^{(s-1)}\biggr)\;,\]
where we have introduced the "edth" operators of Newman and Penrose \cite{NewPen} : 
\bea
\edth_s\sigma^{(s)}&=&
q^{-s+1}{\partial\over\partial\zeta}\bigl(q^s\sigma^{(s)}\bigr)=
q{\partial\sigma^{(s)}\over\partial\zeta}+
s{\partial q\over\partial\zeta}\sigma^{(s)}\nonumber\;,\\
\edth_s^\dagger\sigma^{(s)}&=&
q^{s+1}{\partial\over\partial\zeta^*}\bigl(q^{-s}\sigma^{(s)}\bigr)=
q{\partial\sigma^{(s)}\over\partial\zeta^*}-
s{\partial q\over\partial\zeta^*}\sigma^{(s)}\;.\label{Hil5}\eea
With respect to the scalar product of Pensov scalars :
\be\label{Hil6}
\biggl(\Sigma^{(s)},T^{(s)}\biggr)_s=\int_{S^2}\,\sigma^{(s)*}\,\tau^{(s)}\;\omega\;,\;\ee 
where $\omega=\theta^1\wedge\theta^2={\im\over 2}\theta\wedge\theta^*$ is the invariant volume 
element on $S^2$, 
the operators $\edth_s$ and $\edth_s^\dagger$ are formally anti-adjoint :
\be\label{Hil7}\biggl(\sigma^{(s+1)},\edth_s\tau^{(s)}\biggr)_{s+1}=
\biggl(-\edth_{s+1}^\dagger\sigma^{(s+1)},\tau^{(s)}\biggr)_s\;.\ee
In a previous paper \cite{Migetal1}, the Dirac operator on K\"{a}hler spinors was defined as the
restriction of $-\im(\dif-\delta)$ to the left ideals of the Clifford algebra bundle. 
Now, these ideals are identified with $\Ideal^E_+=\Pe^{(0)}\oplus\Pe^{(+1)}$, with basis 
$\{\Id+\im\omega,\theta\}$, and $\Ideal^E_-=\Pe^{(-1)}\oplus\Pe^{(\bar 0)}$, with basis
$\{\theta^*,\Id-\im\omega\}$.\\ In these bases, the local expressions of the Dirac operators were
given as :
\beann 
\Di^E_+\left(\begin{array}{l}\sigma^{(0)}\\ \sigma^{(+1)}\end{array}\right)&=&
-\im\left(\begin{array}{cc} 0 & \edth^\dagger_{+1}\\ \edth_0 & 0\end{array}\right)
\left(\begin{array}{l}\sigma^{(0)}\\ \sigma^{(+1)} \end{array}\right)\;,\\
\Di^E_-\left(\begin{array}{l}\sigma^{(-1)}\\\sigma^{(\bar 0)}\end{array}\right)&=&
-\im\left(\begin{array}{cc} 0 & \edth^\dagger_0\\\edth_{-1} & 0\end{array}\right)
\left(\begin{array}{l}\sigma^{(-1)}\\ \sigma^{(\bar 0)}\end{array}\right)\;.\eeann
This suggests to define a Pensov spinor field of weight $s$ as a section
\[\Psi_{(s)}=\Sigma^{(s-1/2)}\oplus\Sigma^{(s+1/2)}\]
of the Whitney sum \(\Pe^{(s-1/2)}\oplus\Pe^{(s+1/2)}\), with a Dirac operator 
locally expressed as : 
\be\label{Hil8}
\Di_{(s)}
\left(\begin{array}{l}
\sigma^{(s-1/2)}\\
\sigma^{(s+1/2)}
\end{array}\right)= 
-\im\left(\begin{array}{cc} 0 & \edth_{s+1/2}^\dagger\\
\edth_{s-1/2} & 0
\end{array}\right)
\left(\begin{array}{l}
\sigma^{(s-1/2)}\\
\sigma^{(s+1/2)}
\end{array}\right)\;.\ee 
The usual Dirac spinors on $S^2$ are recovered when $s=0$.\\
With the complex representation of the real Clifford algebra\footnote{
The real Clifford algebra $\Clif(p,q)$ is defined by 
\(\gamma^k\gamma^\ell+\gamma^\ell\gamma^k=2 \eta^{k\ell}\), where the flat metric tensor
$\eta^{k\ell}$ is diagonal with $p$ times $+1$ and 
$q$ times $-1$. 
This entails some differences with other work using the Clifford algebra $\Clif(0,n)$ for 
Riemannian manifolds instead of $\Clif(n,0)$ used here.} \(\Clif(2,0)\) 
\be\label{gammas}
\gamma^1\Rightarrow\left(\begin{array}{cc} 0 & 1\\ 1 & 0\end{array}\right)\;,\;
\gamma^2\Rightarrow\left(\begin{array}{cc} 0 & \im\\ -\im & 0\end{array}\right)\;,\ee
acting on \(\psi_{(s)}=\left(\begin{array}{l}\sigma^{(s-1/2)}\\ \sigma^{(s+1/2)}\end{array}\right)\), 
the Dirac operator can also be written as :
\be\label{Hil9}
\Di_{(s)}\psi_{(s)} =-\im\gamma^k\;\nabla_{k,(s)}^{LC}\psi_{(s)}\;,\ee
where the covariant derivative of the spinor \(\psi_{(s)}\) is given by :
\[\nabla_{k,(s)}^{LC}\psi_{(s)}=
{q\over 2}{\partial\psi_{(s)}\over\partial \xi_k}+{1\over 2}
\Sigma_{ij}^{(s)}\Gamma_k^{\;\;ij}\,\psi_{(s)}\;.\] Here, \(\Sigma_{12}^{(s)}=
\im s\left(\begin{array}{cc}1 & 0\\ 0 & 1\end{array}\right)+
\left(\begin{array}{cc}-\im/2 & 0\\ 0 & +\im/2\end{array}\right)\)
reduces to ${1\over 4}[\gamma_1,\gamma_2]$ for $s=0$.\\
In terms of the "edth" operators, we may write
\beann
\nabla_{(s),+}^{LC}\psi_{(s)}&=&
{1\over 2}
\left(\begin{array}{cc}\edth_{s-1/2}&0\\ 0&\edth_{s+1/2} \end{array}\right)
\psi_{(s)}\;,\nonumber\\
\nabla_{(s),-}^{LC}\psi_{(s)}&=&
{1\over 2}
\left(\begin{array}{cc}\edth^\dagger_{s-1/2}&0\\
0&\edth^\dagger_{s+1/2}\end{array}\right)\psi_{(s)}\;.\eeann
With \(\gamma^{(+)}=\gamma^1+\im\gamma^2=
\left(\begin{array}{cc} 0 & 0\\ 2 & 0\end{array}\right)\) and
\(\gamma^{(-)}=\gamma^1-\im\gamma^2=
\left(\begin{array}{cc} 0 & 2\\ 0 & 0\end{array}\right)\), the Dirac operator of (\ref{Hil8}) is 
now written as 
\[\Di_{(s)}\psi_{(s)}=-\im\left(\gamma^{(+)}\nabla_{(s),+}^{LC}
+\gamma^{(-)}\nabla_{(s),-}^{LC}\right)\psi_{(s)}\;.\]
The transformation law for $s$-Pensov fields\footnote{A Pensov spinor of weight $s$ can be 
interpreted as a usual Dirac spinor on $S^2$, interacting with a Dirac monopole of strength $s$.
Indeed, in the expression of the covariant derivative, the term $\im s\gamma^k\Gamma_k^{\;\;12}$ is
the Clifford representative of the one-form (potential) 
$\mu_s\doteq\im s\theta^k\Gamma_k^{\;\;12}=
{s\over 1+\vert\zeta\vert^2}\left(\zeta^*d\zeta-\zeta d\zeta^*\right)$, which, in
$\left\{H_B\;;\;\cos\theta\neq +1\right\}$, takes the  usual form ${\mu_s}_{\vert B}=\im
s(1-cos\theta)d\phi$.} under a local Zweibein rotation (\ref{Hil2}) is related to the ${\rm
Spin}^c$ structure of the Pensov spinors : 
\[\left(\begin{array}{l}\sigma^{(s-1/2)}\\ \sigma^{(s+1/2)}\end{array}\right)\mapsto
\left(\begin{array}{l}\sigma^{\prime(s-1/2)}\\ \sigma^{\prime(s+1/2)}\end{array}\right)=
\exp\{\alpha\Sigma_{12}^{(s)}\}\left(\begin{array}{l}\sigma^{(s-1/2)}\\
\sigma^{(s+1/2)}\end{array}\right)\;,\]
where
\[\exp\{\alpha\Sigma_{12}^{(s)}\}=
\exp\{\im\;s\alpha\}
\left(\begin{array}{cc}\exp(-\im\alpha/2) & 0\\ 0 & \exp(+\im\alpha/2)\end{array}\right)\;.\]
The Clifford action of $\gamma_3=\im\omega$ yields a grading on the Pensov spinors 
\[\gamma_3\left(\begin{array}{l}
\psi^{(s)}\\ \psi^{(s+1)} \end{array}\right)
=\left(\begin{array}{cc} 1 & 0\\ 0 & -1 \end{array}\right)
\left(\begin{array}{l}\psi^{(s-1/2)}\\ \psi^{(s+1/2)}
\end{array}\right)\;,\;(\gamma_3)^2=\Id\;,\]
such that the Dirac operator (\ref{Hil8}) is odd \[\Di_{(s)}\gamma_3\,+\,\gamma_3\Di_{(s)}\;=\;0\;.\]
According to (\ref{Hil6}), the scalar product of two Pensov spinors is defined as :
\be\label{Hil10}
\langle\Phi_{(s)}\mid\Psi_{(s)}\rangle=
\biggl(\Sigma^{(s-1/2)},T^{(s-1/2)}\biggr)_{s-1/2}\;+\;
\biggl(\Sigma^{(s+1/2)},T^{(s+1/2)}\biggr)_{s+1/2}\;.\ee
The adjointness (\ref{Hil7}) of $-\im\edth_{s-1/2}$ and $-\im\edth^\dagger_{s+1/2}$ implies that 
the Dirac operator is formally self-adjoint with respect to this scalar product. After completion,
\(\Pe^{(s-1/2)}\oplus\Pe^{(s+1/2)}\) becomes a bona fide Hilbert space $\Hi_{(s)}$ on which
$\Di_{(s)}$ acts as a self-adjoint (unbounded) operator. Its spectral resolution is completely
solvable.  Indeed, let $\X=\left(\part \quad \part^{\;*}\right)\;\left(\begin{array}{c}X^+\\
X^-\end{array}\right)$ be a vector field on
$S^2$, then the Lie derivatives of the Zweibein along $\X$ are: 
\bea\label{Hil11}\Lie_{\X}\theta&=&
\biggl({\partial X^+\over\partial\zeta} -{1\over q}({\partial q\over\partial\zeta^*}X^-
+{\partial q\over\partial\zeta}X^+)\biggr)\theta
+{\partial X^+\over\partial\zeta^*}\theta^*\;,\nonumber\\
\Lie_{\X}\theta^*&=&
\biggl({\partial X^-\over\partial\zeta^*} -{1\over q}({\partial q\over\partial\zeta^*}X^-
+{\partial q\over\partial\zeta}X^+)\biggr)\theta^*
+{\partial X^-\over\partial\zeta}\theta
\;.\eea
A vector field $\X$ is said to be a conformal Killing vector field if 
$\Lie_{\X}{\bf g}=\mu\,{\bf g}$, where $\mu$ is a scalar function on $S^2$.
The expression of the Lie derivative (\ref{Hil11}) yields then the (anti-)holomorphic constraints : 
\[{\partial X^+\over\partial\zeta^*}=0\;,\;{\partial X^-\over\partial\zeta}=0\;,\] 
and $\mu$ is given by : 
\[\mu=q^2\biggl({\partial (X^+/q^2)\over\partial\zeta}
+{\partial (X^-/q^2)\over\partial\zeta^*}\biggr)\;.\]
If $\X$ has to be globally defined, its Zweibein components $(2/q)X^+$ and 
$(2/q)X^-$ must be finite when $\mid\zeta\mid\rightarrow\infty$. 
For the standard metric $q=1+\mid \zeta\mid^2$ and this implies that $X^+$, respectively $X^-$, is
a quadratic polynomial in
$\zeta$, respectively $\zeta^*$.\\
There are thus six linearly independent conformal Killing vector fields, three of which are
genuinely Killing,i.e. with $\mu=0$. They are chosen as\footnote{Here we use the $\zeta_B$
coordinates to conform to standard conventions.}
\beann
\im \L_x&\;=\;&{\im\over 2}\biggl((\zeta^2-1)\part -(\zeta^{*2}-1)\part^{\;*}\biggr)\;\\
\im \L_y&\;=\;&{1\over 2}\biggl((\zeta^2+1)\part+(\zeta^{*2}+1)\part^{\;*}\biggr)\;,\\
\im \L_z&\;=\;&\im\biggl(\zeta\part-\zeta^*\part^{\;*}\biggr)\;.\eeann
The other three conformal Killing vector fields (with $\mu\neq 0$) are :
\beann
\im \K_x&\;=\;&{1\over 2}\biggl((\zeta^2-1)\part+(\zeta^{*2}-1)\part^{\;*}\biggr)\;,\\
\im \K_y&\;=\;&{\im\over 2}\biggl((-\zeta^2-1)\part+(\zeta^{*2}+1)\part^{\;*}\biggr)\;,\\
\im \K_z&\;=\;&\biggl(\zeta\part+\zeta^*\part^{\;*}\biggr)\;.\eeann 
As is well known, they form the Lie algebra $sl(2,\Co)$ with its Lie subalgebra
$su(2)$ generated by $\{\im \L_x\;,\;\im \L_y\;,\;\im \L_z\;\}$. Standard angular momentum 
technique tells us that is easier to deal with the complex Killing vectors :
\bea\label{Hil12}
\L_+&=&\L_x+\im\L_y\;=\;\zeta^2\part+\part^{\;*}\;,\nonumber\\
\L_-&=&\L_x-\im\L_y\;=\;-\part-\,\zeta^{*2}\part^{\;*}\;,\nonumber\\
\L_0&=&\L_z\;=\;\zeta\part-\zeta^*\part^{\;*}\;,\eea 
with commutation relations :
\[[\L_0,\L_\pm]=\pm\,\L_\pm\;,\;[\L_+,\L_-]=2\,\L_0\;.\]
The Lie derivatives of $\theta$ with respect to these vector fields are :
\[\Lie_+\theta=\zeta\theta\;,\;\Lie_-\theta=\zeta^*\theta\;,\;\Lie_0\theta=\theta\;.\] 
An infinitesimal transformation of the Zweibein by Killing vectors is a rotation of the 
form (\ref{Hil2}) :
\(\theta\mapsto\tilde\theta=\theta-\im\,\delta\alpha\,\theta=\theta+\delta t\Lie_{\X}\theta\;,\)
and, according to (\ref{Hil11}), we identify 
\[-\im\,\delta\alpha=
\biggl({\partial X^+\over\partial\zeta} -{1\over q}(\zeta X^-+\zeta^* X^+)\biggr)\delta t\;.\]
The transformation of the Pensov basis $\Theta^{(s)}$ is then obtained as :
\[\Theta^{(s)}\mapsto\tilde\Theta^{(s)}=\Theta^{(s)}-\im s\,\delta\alpha\,\Theta^{(s)}=
\Theta^{(s)}+\delta t \Lie_{\X}\Theta^{(s)}\;,\;\hbox{with}\] 
\[\Lie_{\X}\Theta^{(s)}=s\biggl({\partial X^+\over\partial\zeta}-{1\over q}
(\zeta X^-+\zeta^* X^+)\biggr)\Theta^{(s)}\;.\] 
The Lie derivatives of Pensov fields $\Sigma^{(s)}=\sigma^{(s)}\Theta^{(s)}$  along the the 
Killing vectors of (\ref{Hil12}) read :
\beann
\Lie_+^{(s)}\sigma^{(s)}&\;=\;&
(\zeta^2{\partial\over\partial\zeta}
+{\partial\over\partial\zeta^*}+s\zeta)\sigma^{(s)}\;,\\
\Lie_-^{(s)}\sigma^{(s)}&\;=\;&
(-{\partial\over\partial\zeta}
-\zeta^{*2}{\partial\over\partial\zeta^*}+s\zeta^*)\sigma^{(s)}\;,\\
\Lie_0^{(s)}\sigma^{(s)}&\;=\;&
(\zeta{\partial\over\partial\zeta}
-\zeta^*{\partial\over\partial\zeta^*}+s)\sigma^{(s)}\;.\eeann 
These Lie derivatives yield a representation of $su(2)$ on $s$-Pensov fields :
\[ [\Lie_0^{(s)},\Lie_\pm^{(s)}]=\pm\Lie_\pm^{(s)}\;,
\;[\Lie_+^{(s)},\Lie_-^{(s)}]=2\Lie_0^{(s)}\;.\]
The Casimir operator is given by :
\[(\vec\Lie^{(s)})^2={1\over 2}
\biggl(\Lie_+^{(s)}\Lie_-^{(s)}\,+\,\Lie_-^{(s)}\Lie_+^{(s)}\biggr)+(\Lie_0^{(s)})^2=
-q^2{\partial^2\over\partial\zeta\partial\zeta^*}+s\,q\,\Lie_0^{(s)}\;.\]
Straightforward angular momentum algebra yields the monopole harmonics of Wu and Yang \cite{W-Y},
solutions of 
\be\label{Hil13} (\vec\Lie^{(s)})^2\;\Y^s_{j,m}=j(j+1)\;\Y^s_{j,m}\;,\;
\Lie_0^{(s)}\;\Y^s_{j,m}=m\;\Y^s_{j,m}\;,\ee
where $j=\mid s\mid,\mid s\mid+1,\cdots$ and $m=-j,-j+1,\cdots,j-1,j$.\\ 
They can be written in terms of Jacobi functions and, using some appropriate Olinde-Rodrigues formulae, they are obtained 
as :
\bea\label{Har}
\Y^s_{j,m}&=&{(-1)^{j-s}\over 2^j}\sqrt{2j+1\over 4\pi}\sqrt{(j+m)!\over(j+s)!(j-s)!(j-m)!}
\exp(\im(m-s)\varphi)\nonumber\\ &&(1-z)^{-{m-s\over 2}}(1+z)^{-{m+s\over 2}}
\biggl({d\over dz}\biggr)^{j-m}\biggl((1-z)^{j-s}(1+z)^{j+s}\biggr)\;\nonumber\\
&=&{(-1)^{j+m}\over 2^j}\sqrt{2j+1\over 4\pi}\sqrt{(j-m)!\over(j+s)!(j-s)!(j+m)!}
\exp(\im(m-s)\varphi)\nonumber\\ &&(1-z)^{{m-s\over 2}}(1+z)^{{m+s\over 2}}
\biggl({d\over dz}\biggr)^{j+m}\biggl((1-z)^{j+s}(1+z)^{j-s}\biggr)\;,\eea 
where $\varphi$ is the azimuthal angle and $z=\cos(\theta)$, the cosine of the polar angle.\\   
The Lie derivative of Pensov fields along one of the vector fields 
\({\bf L}_\pm,{\bf L}_0\) "commutes" with the edth operators of (\ref{Hil5}) in the sense that\footnote{This is quite expected since the edth operators are constructed  from the Levi-Civita connection which is metric compatible and the Killing vector fields conserve this metric.}:
\[\edth_s\;\vec\Lie^{(s)}=\vec\Lie^{(s+1)}\;\edth_s\;,\;
\edth_{s+1}^\dagger\;\vec\Lie^{(s+1)}=\vec\Lie^{(s)}\;\edth_{s+1}^\dagger\;.\] 
With the choice of phases in (\ref{Har}), one has
\beann
\edth_s\Y^s_{j,m}&=&-\sqrt{(j-s)(j+s+1)}\;\Y^{s+1}_{j,m}\;,\\
\edth_{s+1}^\dagger\Y^{s+1}_{j,m}&=&
+\sqrt{(j-s)(j+s+1)}\;\Y^s_{j,m}\;.\eeann 
On Pensov spinors of $\Hi_{(s)}$, one defines the "total angular momentum" as :
\[\vec\Lie_{tot}^{(s)}\left(\begin{array}{l} \sigma^{(s-1/2)}\\ \sigma^{(s+1/2)}\end{array}\right)
=\left(\begin{array}{cc}
\vec\Lie^{(s-1/2)} & 0\\ 0 & \vec\Lie^{(s+1/2)}\end{array}\right)
\left(\begin{array}{l} \sigma^{(s-1/2)}\\ \sigma^{(s+1/2)}\end{array}\right)
\;.\]
It commutes with the Dirac operator : \(\Di_{(s)}\vec\Lie_{tot}^{(s)}\;=\;\vec\Lie_{tot}^{(s)}\Di_{(s)}\). \\
From the product of edth operators 
\beann -\edth_{s+1/2}^\dagger\;\edth_{s-1/2}&=&(\vec\Lie^{(s-1/2)})^2-\,(s^2-1/4)\;,\\
-\edth_{s-1/2}\;\edth_{s+1/2}^\dagger&=&(\vec\Lie^{(s+1/2)})^2-\,(s^2-1/4)\;,\eeann 
a Lichnerowicz type formula follows immediately :
\[{\Di_{(s)}}^2=(\vec\Lie_{tot}^{(s)})^2-\;(s^2-1/4)\,\Id\;.\]
Using (\ref{Hil13}), the eigenvalues of the Dirac operator are found to be :
\[\Di_{(s)}\psi_{(s),j,m}^{(\pm)}=
\pm\sqrt{(j+1/2)^2-s^2}\;\psi_{(s),j,m}^{(\pm)}\;,\]
with 
\[\psi_{(s),j,m}^{(+)}=
\left(\begin{array}{c} {1\over\sqrt{2}}\Y^{s-1/2}_{j,m}\\ {\im\over\sqrt{2}}\Y^{s+1/2}_{j,m}
\end{array}\right)\;,\;\;
\psi_{(s),j,m}^{(-)}=
\gamma_3\,\psi^{(+)}_{(s),j,m}=
\left(\begin{array}{c} {1\over\sqrt{2}}\Y^{s-1/2}_{j,m}\\ -{\im\over\sqrt{2}}\Y^{s+1/2}_{j,m}
\end{array}\right)\;.\] 
In particular it follows that, for Dirac spinors i.e. $s=0$, and only in this case, there are no zero eigenvalues. When $s\not=0$, zero is an eigenvalue, $2\vert s\vert$ times degenerate with eigenspinors 
\[
\left\{\begin{array}{ll}
\left(\begin{array}{c} \Y^{s-1/2}_{s-1/2,m}\\ 0 \end{array}\right) &
\hbox{ for positive values of s}\;,\\
&\\
\left(\begin{array}{c} 0 \\ \Y^{s+1/2}_{-s-1/2,m}\end{array}\right) &
\hbox{ if s is negative.}\end{array}\right.\]
%%%%%% 
%%%%%%
\newpage
%%%%%%
%%%%%%
\section{The projective modules over $\Al$}\label{Module}
\setcounter{equation}{0} 
Through the Gel'fand-Na\u{\i}mark construction, the topology of $M=S^2\times\{a,b\}$ is encoded 
in the complex $C^*$-algebra of continuous complex-valued functions on $M=S^2\times\{a,b\}$.
However, in order to get a fruitful use of a differential structure, we have to restrict this
$C^*$-algebra to its dense subalgebra of smooth functions. This proviso made, let
$\{f,g,\cdots\}$ denote elements of $\Al=\Fu(M)$ and let the value of $f$ at a point\footnote{In
this section points of $S^2$ are denoted by $x,y,\cdots$, while $\alpha,\beta,\cdots$ will assume
values in the two-point space $\{a,b\}$. Points of $M=S^2\times\{a,b\}$ are thus written as 
$p=\{x,\alpha\},q=\{y,\beta\}$, etc. and the value of a function $F$ at $(p,q,\cdots)$ will also be expressed as 
$F_{\alpha,\beta,\cdots}(x,y,\cdots)$.} $p=\{x,\alpha\}\in M$, be written as $f(p)=f_\alpha(x)$.
The vectors of the free right $\Al$-module of rank two, identified with $\Al^2$, are of the form 
$X=\sum_{\;i=1,2}E_i\;f^i$, where $f^i\in\Al$ and $\{E_i\,;i=1,2\}$ is a basis of $\Al^2$. 
Let $\Omega^\bullet(\Al)=\sum_{\;k=0}^{\infty}\Omega^{(k)}(\Al)$ denote the universal differential envelope of $\Al$. Elements of $\Omega^{(k)}(\Al)$ can be  realised, see e.g. \cite{Coq}, as functions on the Cartesian product of $(k+1)$ copies of $M$, vanishing on neighbouring diagonals, i.e.
\(F(p_0,p_1,\cdots,p_k)=0\) if, for some $i$, $p_i=p_{i+1}$.\\
The product in $\Omega^\bullet(\Al)$ is obtained by concatenation, e.g. if $F\in\Omega^{(1)}(\Al)$ and 
$G\in\Omega^{(2)}(\Al)$ then their product $F\cdot G\in \Omega^{(3)}(\Al)$ is represented by 
\[(F\cdot G)(p_0,p_1,p_2,p_3)=F(p_0,p_1)\;G(p_1,p_2,p_3)\;.\]
The differential $\dif$ acts on $f\in\Omega^{(0)}(\Al)$ and on $F\in\Omega^{(1)}(\Al)$ as follows :
\[(\dif f)(p_0,p_1)=f(p_1)-f(p_0)\;,\]  
\[(\dif F)(p_0,p_1,p_2)=F(p_1,p_2)-F(p_0,p_2)+F(p_0,p_1)\;.\]
The involution\footnote{Note that $\dif(f^\dagger)=-(\dif f)^\dagger\;,f\in\Al$ and, more generally, if $F\in\Omega^{(k)}(\Al)$, then $\dif(F^\dagger)= (-1)^{k+1}(\dif F)^\dagger$.}, defined in $\Al$ by $(f^\dagger)(p)=\Bigl(f(p)\Bigr)^*$, extends to $\Omega^{(k)}(\Al)$ as
\[(F^\dagger)(p_1,p_2,\cdots)=\Bigl(F(\cdots,p_2,p_1)\Bigr)^*\;.\]
A (universal) connection on $\Al^2$ is given, in the basis $\{E_i\;;i=1,2\}$, by an 
$\Omega^{(1)}(\Al)$-valued $2\times 2$ matrix $\fatl\omega\fatr^i_{\;k}$. 
It acts on $X=E_i\;f^i$ as :
\be\label{PMod1}
\nabla_{free}(X)=E_i\otimes_\Al\Bigl(\dif f^i+\fatl\omega\fatr^i_{\;k} f^k\Bigr)\;.\ee 
Let $X=E_i\;f^i$ and $Y=E_i\;g^i$ be two vectors of $\Al^2$, then the standard hermitian product 
with values in $\Al$ is given by :
\[{\bf h}(X,Y)=\sum_{\;i,j}\;(f^i)^\dagger\;\delta_{\overline{\i}j}g^j=
\sum_{\;i}\;(f^i)^\dagger\;g^i\;\]
It extends as $\Omega^{\bullet}(\Al)$-valued on 
$\Bigl(\Al^2\otimes_\Al\Omega^{\bullet}(\Al)\Bigr)\times
\Bigl(\Al^2\otimes_\Al\Omega^{\bullet}(\Al)\Bigr)$ as 
\[{\bf h}(X\otimes F,Y\otimes G)= F^\dagger{\bf h}(X,Y) G\;.\]
The connection is hermitian if \(\dif\Bigl({\bf h}(X,Y)\Bigr)=
{\bf h}(X,\nabla_{free} Y)-{\bf h}(\nabla_{free} X,Y)\). \\
For the product above, this yields 
\(\fatl\omega\fatr^i_{\;j}=
\delta^{i\overline{k}}{\fatl\omega\fatr^\ell_{\;k}}^\dagger\delta_{\overline{\ell}j}\).\\ 
The representation of $\omega$ by functions on $M\times M$ is given by :
\[\fatl\omega\fatr\Rightarrow\left(
\begin{array}{cc}
K_{\alpha\beta}(x,y) & S_{\alpha\beta}(x,y)\\
T_{\alpha\beta}(x,y) & L_{\alpha\beta}(x,y)\end{array}\right)\] 
and the hermiticity condition reads :
\bea\label{PMod2}
K_{\alpha\beta}(x,y)&=&\Bigl(K_{\beta\alpha}(y,x)\Bigr)^*\;,\nonumber\\
L_{\alpha\beta}(x,y)&=&\Bigl(L_{\beta\alpha}(y,x)\Bigr)^*\:,\nonumber\\
T_{\alpha\beta}(x,y)&=&\Bigl(S_{\beta\alpha}(y,x)\Bigr)^*\;.\eea
The action of the connection (\ref{PMod1}) is represented by :
\[(\nabla_{free} X)^i_{\alpha\beta}(x,y)=
f^i_\beta(y)-f^i_\alpha(x)+{\fatl\omega\fatr^i}_{k,\alpha\beta}(x,y)\;f_\beta^k(y)\;.\] 
A projective module is defined by an endomorphism $\Proj$ of
$\Al^2$ which is idempotent, $\Proj^2=\Proj$, and hermitian, $\Proj^\dagger=\Proj$, where 
the adjoint ${\bf A}^\dagger$ of an endomorphism ${\bf A}$ is defined by ${\bf h}(X,{\bf A}^\dagger
Y)={\bf h}({\bf A}X,Y)$.  In the basis $\{E_i\}$, the projector is given by a $2\times 2$ matrix
${(P)^i}_j$ with entries in $\Al$ and is  represented by ${(P)^i}_{j,\alpha}(x)$. The projective
module $\Mod$ is defined as the image of $\Proj$ :
\[\Mod=\Bigl\{\Proj X\;\vert\;X\in \Al^2\Bigr\}=\Bigl\{X\in\Al^2\;\vert\;\Proj X=X\Bigr\}\;.\]
The hermiticity of the projector guarantees that ${\bf h}$, restricted to $\Mod$, defines a 
hermitian product in $\Mod$.\\
In the Connes-Lott model \cite{C-L}, the projectors are of the form
\[{\fatl P\fatr^i}_{j,a}(x)={\delta^i}_j\;\hbox{and}\;
{\fatl P\fatr^i}_{j,b}(x)= {1\over 2}\Bigl(\Id+\vec n(x).\vec\sigma\Bigr)^i_{\;j}\;,\]
where $\vec\sigma$ are the Pauli matrices and $\vec n(x)$ is a real unit vector, mapping 
$S^2\mapsto S^2$ so that the projectors are classified by $\pi_2(S^2)=\Ent$. Furthermore, 
since $\pi_2(U(2))=\pi_2(SU(2))=\pi_2(S^3)=\{\Id\}$, projectors, belonging to
different homoptopy classes, cannot be unitarily equivalent.\\
The target sphere $S^2$ also has two coordinate charts $H_B^{target}$ and $H_A^{target}$.\\
In these charts, the projector $\fatl P_b\fatr$ can be written as 
\(\fatl P_b^B\fatr=\vert\nu_B\rangle \langle\nu_B\vert\), respectively \(\fatl
P_b^A\fatr=\vert\nu_A\rangle \langle\nu_A\vert\), where $\nu_B$, respectively $\nu_A$, is the
complex coordinate of $\vec n$ in $H_B^{target}$, respectively $H_A^{target}$. We have used the
Dirac ket- and bra-notation : 
\[\vert\nu_B\rangle={1\over\sqrt{1+\vert\nu_B\vert^2}}
\left(\begin{array}{c} 1\\ \nu_B \end{array}\right)\quad,\quad
\langle \nu_B\vert={1\over\sqrt{1+\vert\nu_B\vert^2}}
\Bigl(1\;\quad\;\nu_B^*\Bigr)\;,\]
\[\vert\nu_A\rangle={1\over\sqrt{1+\vert\nu_A\vert^2}}
\left(\begin{array}{c} -\nu_A \\ 1\end{array}\right)\quad,\quad
\langle \nu_A\vert={1\over\sqrt{1+\vert\nu_A\vert^2}}
\Bigl(-\nu_A^* \;\quad\; 1\Bigr)\;.\]
An element  $X$ of $\Al^2$ is represented by the column matrix 
\( X\Rightarrow
\left(\begin{array}{c}
\vert f_a(x)\rangle \\ \vert f_b(x)\rangle\end{array}\right)\), where
\(\vert f_a(x)\rangle=\left(\begin{array}{c} f_a^1(x) \\ f_a^2(x)\end{array}\right)\) and 
\(\vert f_b(x)\rangle=\left(\begin{array}{c} f_b^1(x) \\ f_b^2(x)\end{array}\right)\). 
It belongs to $\Mod$ if $\Proj X = X$, which yields no restriction on \(\vert f_a(x)\rangle\)
but there is one on \(\vert f_b(x)\rangle\).
In $H_B^{target}$ it is expressed as $\vert f_b\rangle=\vert\nu_B\rangle\;f_b^B$, where
\[f_b^B=\langle\nu_B\vert f_b\rangle
={1\over\sqrt{1+\vert\nu_B\vert^2}}\bigl(f_b^1+\nu_B^*f_b^2\bigr)\;.\] 
In the same way, in $H_A^{target}$ one has $\vert f_b\rangle=\vert\nu_A\rangle\;f_b^A\;,$ where
\[f_b^A=\langle\nu_A\vert f_b\rangle
={1\over\sqrt{1+\vert\nu_A\vert^2}}\bigl(-\nu_A^*f_b^1+f_b^2\bigr)\;.\] 
As representatives of the homotopy class $[n]\in\pi_2(S^2)\equiv\Ent$, we choose a 
mapping $\vec n$ transforming $H_B$, respectively $H_A$ of the range $S^2$, into $H_B^{target}$,
respectively $H_A^{target}$ of the target $S^2$.  Such a choice is\footnote{Note that this choice
is different from the one in previous work \cite{Migetal1}.}
\be\label{PMod3}
\nu_B(x)=\biggl({\zeta_B\over\zeta_B^*}\biggr)^{n-1\over 2}\zeta_B\;,\;
\nu_A(x)=\biggl({\zeta_A\over\zeta_A^*}\biggr)^{n-1\over 2}\zeta_A\;;\;n\in\Ent\;,\ee 
where $\zeta_B\;,\;\zeta_A$ are the complex coordinates of $x\in S^2$.\\ 
In the overlap $H_A\cap H_B$, with transition function $c_{AB}$ given by (\ref{Hil3}), we have:
\[ f_b^A(x) =\Bigl(c_{AB}(x)\Bigr)^{n/2}\;f_b^B(x)\]
and this tells us that $f_b(x)$ is a Pensov scalar of weight $n/2$.\\ 
In the rest of this paper we shall omit the $A$ and $B$ labels except when relating quantities 
in $H_A$ with those in $H_B$ in the overlap $H_A\cap H_B$.\\
An element of $\Mod$ is thus represented by :
\(X\Rightarrow
\left(\begin{array}{c}
\vert f_a\rangle \\\vert \nu\rangle f_b\end{array}\right)\) and its scalar product with 
\(Y\Rightarrow
\left(\begin{array}{c}
\vert g_a\rangle \\ \vert \nu\rangle g_b \end{array}\right)\) is given by :
\beann
\Bigl({\bf h}_\Proj(X,Y)\Bigr)_a(x)&=
&\Bigl(f_a^1(x)\Bigr)^*g_a^1(x)+\Bigl(f_a^2(x)\Bigr)^*g_a^2(x)\;,\\
\Bigl({\bf h}_\Proj(X,Y)\Bigr)_b(x)&=
& \Bigl(f_b(x)\Bigr)^*g_b(x)\;.\eeann 
An active gauge transformation in the free module $\Al^2$ is given by a unitary $2\times 2$
matrix $\U$ with values in $\Al$. It retricts to a gauge transformation in $\Mod$ when it
commutes with $\Proj$ : $\Proj\U=\U\Proj$.\\ 
An element $X\in \Mod$ transforms as $X\mapsto \U X$ given by :
\bea\label{PMod4}
\vert\Bigl(\U X\Bigr)_a(x)\rangle &=& \fatl\U_a(x)\fatr\;\vert f_a(x)\rangle\;,\nonumber\\
\vert\Bigl(\U X\Bigr)_b(x)\rangle &=& \vert\nu(x)\rangle\bu_b(x)\;f_b(x)\;,\eea 
where $\fatl\U_a(x)\fatr\in U(2)$ and $\bu_b(x)\in U(1)$.\\ 
The connection in the free module (\ref{PMod1}) induces a connection in the projective
module $\Mod$ given by $\nabla X=\Proj\;\nabla_{free} X$ where $X\in\Mod$.\\ 
It is represented by
\be\label{PMod5} \vert(\nabla X)_{\alpha\beta}(x,y)\rangle=
\fatl P_\alpha(x)\fatr (\vert f_\beta(y)\rangle
- \vert f_\alpha(x)\rangle)+\fatl A_{\alpha\beta}(x,y)\fatr\;\vert f_\beta(y)\rangle\;.\ee 
The $2\times 2$ matrices $\fatl A_{\alpha\beta}(x,y)\fatr$ are given
by :
\bea\label{PMod6} 
\fatl A_{aa}(x,y)\fatr&=&\fatl\omega_{aa}(x,y)\fatr\;,\nonumber\\
\fatl A_{ab}(x,y)\fatr&=&\vert \Phi_{ab}(x,y)\rangle\langle\nu(y)\vert\;,\nonumber\\
\fatl A_{ba}(x,y)\fatr&=&\vert\nu(x)\rangle\,\langle\Phi_{ba}(x,y)\vert\;,\nonumber\\
\fatl A_{bb}(x,y)\fatr&=&\vert\nu(x)\rangle\,\omega_b(x,y)\langle\nu(y)\vert\;,\eea 
where we have introduced the $\Omega^{(1)}(\Al)$-valued ket- and bra- vectors :
\bea\label{PMod7}
\vert\Phi_{ab}(x,y)\rangle&=&\fatl\omega_{ab}(x,y)\fatr\vert\nu(y)\rangle\;,\nonumber\\
\langle\Phi_{ba}(x,y)\vert&=&\langle\nu(x)\vert\fatl\omega_{ba}(x,y)\,\fatr\;,\eea 
and the universal one-form :
\be\label{PMod8}
\omega_b(x,y)=\langle\nu(x)\vert\fatl\omega_{bb}(x,y)\fatr\vert\nu(y)\rangle\;.\ee 
The hermiticity condition (\ref{PMod2}) yields ;
\bea\label{PMod9}
\fatl\omega_{aa}(x,y)\fatr^+&=&\fatl\omega_{aa}(y,x)\fatr\;,\nonumber\\
(\omega_b(x,y))^*&=&\omega_b(y,x)\;,\nonumber\\
\vert\Phi_{ab}(x,y)\rangle^+&=&\langle\Phi_{ba}(y,x)\vert\;.\eea
In $H_B\cap H_A$ :
\bea\label{PMod10}
\vert\Phi^A_{ab}(x,y)\rangle
&=&\vert\Phi^B_{ab}(x,y)\rangle\,\Bigl(c_{AB}(y)\Bigr)^{-n/2}\;,\nonumber\\
\langle\Phi^A_{ba}(x,y)\vert
&=&\Bigl(c_{AB}(x)\Bigr)^{n/2}\,\langle\Phi^B_{ba}(x,y)\vert\;,\nonumber\\
\omega^A_b(x,y)&=&\Bigl(c_{AB}(x)\Bigr)^{n/2}\,\omega^B_b(x,y)\,\Bigl(c_{AB}(y)\Bigr)^{-n/2}\;.\eea
The action of $\nabla$ on $X\in\Mod$ is obtained 
using (\ref{PMod6}), (\ref{PMod7}) and (\ref{PMod8}) :
%%%%%
\bea\label{PMod11} 
\vert(\nabla X)_{aa}(x,y)\rangle&=&
\vert f_a(y)\rangle-\vert f_a(x)\rangle
+\fatl\omega_{aa}(x,y)\fatr\,\vert f_a(y)\rangle\;,\nonumber\\
\vert(\nabla X)_{ab}(x,y)\rangle&=&
\vert H_{ab}(x,y)\rangle \; f_b(y) - \vert f_a(x)\rangle\;,\nonumber\\
\vert(\nabla X)_{ba}(x,y)\rangle&=&
\vert\nu(x)\rangle\,\biggl[\langle H_{ba}(x,y)\vert f_a(y)\rangle -f_b(x)\biggr]\;,\nonumber\\
\vert(\nabla X)_{bb}(x,y)\rangle&=&
\vert\nu(x)\rangle\,\Bigl[f_b(y)-f_b(x)\nonumber\\
&&\quad\quad\quad\quad +(\omega_b(x,y)+ m_b(x,y))f_b(y)\Bigr]\;,
\eea 
where 
\bea\label{PMod12} 
\vert H_{ab}(x,y)\rangle&=&\vert\Phi_{ab}(x,y)\rangle+\vert\nu(y)\rangle\;,\nonumber\\
\langle H_{ba}(x,y)\vert&=&\langle\Phi_{ba}(x,y)\vert+\langle\nu(x)\vert\;.\eea 
and the "monopole" connection $m_b(x,y)$ appears as :
\be\label{PMod13}
m_b(x,y)=\langle\nu(x)\vert\nu(y)\rangle\,-1\;.\ee
As seen from (\ref{PMod10}) and (\ref{PMod12}), the off-diagonal connections 
$\vert H_{ab}(x,y)\rangle$, $\langle H_{ba}(x,y)\vert$ and also $\omega_b(x,y)$ transform 
homogeneously from $H_B$ to $H_A$ but $m_b(x,y)$ transforms with the expected inhomogeneous term :
\bea\label{PMod14}
m^A_b(x,y)&=&
(c_{AB}(x))^{n/2}\,m^B_b(x,y)(c_{AB}(y))^{-n/2}\nonumber\\
&& +(c_{AB}(x))^{n/2}[(c_{AB}(y))^{-n/2}-(c_{AB}(x))^{-n/2}]\;.\eea 
In terms of abstract universal differential one forms, (\ref{PMod13}) and (\ref{PMod14}) read :
\beann
m_b&=&{1\over\sqrt{1+\vert\nu\vert^2}}\Bigl(\nu^*\dif\nu - \sqrt{1+\vert\nu\vert^2}
\dif \sqrt{1+\vert\nu\vert^2}\Bigr){1\over\sqrt{1+\vert\nu\vert^2}}\;,\\  
m_b^A&=&(c_{AB})^{n/2}\,m_b^B\,(c_{AB})^{-n/2}+(c_{AB})^{n/2}\dif(c_{AB})^{-n/2}\;.\eeann 
The curvature of the connection is defined by :
\(\vert{\nabla}^2X\rangle = \fatl R\fatr\vert X\rangle\). It is a right-module 
homomorphism $\Mod\rightarrow\Mod\otimes_\Al\Omega^{(2)}(\Al)$ given
in the basis$\{E_i\}$ by the $2\times 2$ matrix with values in $\Omega^{(2)}(\Al)$ : 
\[\fatl R \fatr= \fatl P\fatr \Bigl(\dif\fatl A\fatr\Bigr) \fatl P\fatr+ \fatl A\fatr^2 + 
\fatl P\fatr \Bigl(\dif \fatl P\fatr\Bigr)\Bigl( \dif \fatl P\fatr\Bigr) \fatl P\fatr\;,\]
or, within the used realisation, by
\bea\label{PMod15}
\lefteqn{\fatl R_{\alpha\beta\gamma}(x,y,z)\fatr=}\hspace{5ex}\nonumber \\
&&\fatl P_\alpha(x)\fatr
\Bigl(\fatl A_{\beta\gamma}(y,z)\fatr-\fatl A_{\alpha\gamma}(x,z)\fatr+
\fatl A_{\alpha\beta}(x,y)\fatr\Bigr)\fatl P_\beta(z)\fatr\nonumber\\
&& + \fatl A_{\alpha\beta}(x,y)\fatr\,\fatl A_{\beta\gamma}(y,z)\fatr\nonumber\\
&& +\fatl P_\alpha(x)\fatr \Bigl(\fatl P_\beta(y)\fatr-\fatl P_\alpha(x)\fatr\Bigr)
\Bigl(\fatl P_\gamma(z)\fatr-\fatl P_\beta(y)\fatr\Bigr)\fatl P_\gamma(z)\fatr\;.\nonumber\\
&&\eea
A connection $\nabla$ compatible with the hermitian structure in $\Mod$ implies in a 
self-adjoint curvature :
\be\label{PMod16}
{R^i}_j=\delta^{i\bar\ell} ({R^k}_\ell)^+\delta_{\bar k j}\;.\ee
%%%%%%%%%%%%
Let the connection $\nabla$ be extended to $\Mod\otimes_\Al\Omega^\bullet(\Al)$ by 
\[\nabla\Bigl(X\otimes_\Al F\Bigr)=(\nabla X)F+X\otimes_\Al \dif F\;,\]
then $\nabla^2$ becomes an endomorphism of the right $\Omega^\bullet(\Al)$-module 
$\Mod\otimes_\Al\Omega^\bullet(\Al)$.\\
The active gauge transformation (\ref{PMod4}) acts on the right on the space of connections 
as $\nabla\mapsto\nabla^\U=\U^{-1}\circ\nabla\circ\U$. The action of $\nabla^\U$ on $X$ is 
given by a similar expression as in (\ref{PMod5}) with the matrices $\fatl A\fatr$ replaced 
by $\fatl A^\U\fatr=\fatl \U\fatr^{-1}\fatl A\fatr \fatl\U\fatr+
\fatl P\fatr\fatl\U^{-1}\fatr(\dif\fatl\U\fatr)\fatl P\fatr$ and (\ref{PMod11}) becomes 
\beann 
\vert(\nabla^\U X)_{aa}(x,y)\rangle&=&
\vert f_a(y)\rangle - \vert f_a(x)\rangle + \fatl\omega_{aa}^\U(x,y)\fatr\,\vert f_a(y)\rangle\;,\\
\vert(\nabla^\U X)_{ab}(x,y)\rangle&=&
\vert H_{ab}^{\U}(x,y)\rangle f_b(y) - \vert f_a(x)\rangle\;,\\
\vert(\nabla^\U X)_{ba}(x,y)\rangle&=&
\vert\nu(x)\rangle\,\biggl[\langle H_{ba}^{\U}(x,y)\vert f_a(y)\rangle-f_b(x)\biggr]\;,\\
\vert(\nabla^\U X)_{bb}(x,y)\rangle&=&
\vert\nu(x)\rangle\,\Bigl[f_b(y)-f_b(x)\nonumber\\
&&\quad\quad\quad\quad+(\omega_b^{\U}(x,y)+m_b^{\U}(x,y))f_b(y)\Bigr]\;.
\eeann 
with
\bea\label{PMod17}
\fatl\omega_{aa}^\U(x,y)\fatr&=&\fatl\U_a(x)\fatr^{-1}
\fatl\omega_{aa}(x,y)\fatr\fatl\U_a(y)\fatr\nonumber\\
&&+\fatl(\U_a(x)\fatr^{-1}\Bigl(\fatl\U_a(y)\fatr-\fatl\U_a(x)\fatr\Bigr)\;,\nonumber\\
\vert H_{ab}^{\U}(x,y)\rangle&=&
\fatl\U_a(x)\fatr^{-1}\;\vert H_{ab}(x,y)\rangle\;\bu_b(y)\;,\nonumber\\
\langle H_{ba}^{\U}(x,y)\vert&=&
(\bu_b(x))^{-1}\;\langle H_{ba}(x,y)\vert\;\fatl\U_a(y)\fatr\;,\nonumber\\
m_b^{\U}(x,y)&=&
(\bu_b(x))^{-1}\;m_b(x,y)\;\bu_b(y)\nonumber\\
\omega_b^{\U}(x,y)&=&
(\bu_b(x))^{-1}\;\omega_b(x,y)\;\bu_b(y)\nonumber\\
&&+(\bu_b(x))^{-1}\Bigl(\bu_b(y)-\bu_b(x)\Bigr)\;.\eea
It is thus seen that $\vert H_{ab}(x,y)\rangle\,,\,\langle H_{ba}(x,y)\vert$ and
the monopole connection (\ref{PMod13}) $m_b(x,y)$ transform homogeneously under an 
active gauge transformation, while $\fatl\omega_{aa}(x,y)\fatr$ and $\omega_b(x,y)$ have 
the expected inhomogeneous terms $\U^{-1}\dif\U$, $\bu^{-1}\dif\bu$.
%%%%%
\newpage
%%%%%
\section{The spectral triple $\Bigl\{\Al,\Hi,\Dirac,\Gamma\Bigr\}$}\label{Triple}
\setcounter{equation}{0}
An (even) spectral triple,$\{\Al,\Hi,\Dirac,\chi\}$, as defined by Connes \cite{Con1}, is given by
a $C^\star$ algebra $\Al$ and a Hilbert space $\Hi$, graded by $\chi$, with 
\(\pi:\Al\mapsto{\cal B}(\Hi): f\mapsto \pi(f)\), a faithful, ${\rm Ker}(\pi)=0$, and even,
 \(\Bigl[\chi,\pi(f)\Bigr]=0\), $\star$-representation, \(\pi(f^*)=\Bigl(\pi(f)\bigr)^+\), of $\Al$
acts as bounded operators.\\ Furthermore, on $\Hi$, there is a self-adjoint Dirac operator
$\Dirac$, which is odd, \(\Dirac\chi+\chi\Dirac=0$, and such that $(\Dirac-\lambda)^{-1}$ is
compact for $\lambda \not\in \Re$. It should be a first order operator in the sense that
\(\Bigl[[\Dirac,\pi(f)],\pi(g)\Bigr]=0\;,\;f,g\in\Al\).\\ Here the algebra is
$\Al=\Fu(S^2\times\{a,b\};\Co)$ and as Hilbert space we take 
\[\Hi=\Hi_{(s)}\otimes\Hi_{dis}\;,\] 
the tensor product of the Hilbert space $\Hi_{(s)}$ of Pensov spinors with a finite Hilbert space 
\(\Hi_{dis}=\Bigl(\Co^{N_a}\oplus\Co^{N_b}\Bigr)\) where
$N_a,N_b$ are natural numbers giving the number of generations in each chirality sector.\\ 
\(\Hi_{dis}\) is endowed with a grading operator: 
\be\label{TRIP0}\chi_{dis}=\left(\begin{array}{cc}
\Id_a & 0 \\ 0 & -\Id_b\end{array}\right)\;,\ee
where $\Id_\alpha\;,\;\alpha=a,b$ is the $N_\alpha\times N_\alpha$ unit matrix.
On $\Hi_{dis}$ there a finite Dirac operator $\Dirac_{dis}$ represented by a hermitian matrix 
odd with respect to $\chi_{dis}$ :
\be\label{TRIP1}
\Di_{dis}=\left(\begin{array}{cc}0 & M^+\\ M & 0\end{array}\right)\;,\ee
where $M$ is a $N_b\times N_a$ matrix describing the phenomenology of the 
masses.
The total grading in \(\Hi\) is given by 
\be\label{TRIP01}\chi=\gamma_3\otimes\chi_{dis}=\left(\begin{array}{cc}
\gamma_3\otimes\Id_a & 0 \\ 0 & 
-\gamma_3\otimes\Id_b\end{array}\right)\;,\ee
The Dirac operator in $\Hi$ is obtained from (\ref{Hil9}) and (\ref{TRIP1}) as ;
\be\label{TRIP2}
\Dirac=\Dirac_{(s)}\otimes\Id_{a+b}+\gamma_3\otimes\Dirac_{dis}\;.\ee
It is odd with respect to the total grading $\chi$. The grading operator $\gamma_3$ has 
been introduced in (\ref{TRIP2}) so that the square of the total Dirac operator reads
\[\Dirac^2
={\Dirac_{(s)}}^2\otimes\Id_{a+b}+\Id\otimes{\Dirac_{dis}}^2\;.\]
An element $\BPsi$ of $\Hi$ is represented as
\be\label{TRIP3}
\left(\begin{array}{c}\psi_{(s),a}(x)\\ \psi_{(s),b}(x)\end{array}\right)\;,\ee
where each $\psi_{(s),\;\alpha}(x),(\alpha=a,b)$, is a Pensov spinor with 
$N_\alpha$ "generation" indices\footnote{These "generation" indices are not written down
explicitely and the $(s)$ subscript, fixed once for all, will also be omitted in this section.}.
The scalar product is the obvious extension of (\ref{Hil10}) :
\be\label{TRIP4}
(\BPsi;\BPhi)=\int_{S^2}\biggl[(\psi_a)^+\phi_a\,+\,(\psi_b)^+\phi_b\biggr]\omega\;.\ee
An element $f\in\Al$ is represented as:
\be\label{TRIP5}
\pi(f)\;\left(\begin{array}{c}\psi_a(x)\\ \psi_b(x)\end{array}\right)=
\left(\begin{array}{cc}
f_a(x)\Id_c\otimes\Id_a & 0 \\
0 & f_b(x)\Id_c\otimes\Id_b
\end{array}\right)\;\left(\begin{array}{c}\psi_a(x)\\ \psi_b(x)\end{array}\right)\;,\ee
where $\Id_c$ is the $2\times 2$ unit matrix in the Clifford algebra.
Acting on vectors of the form (\ref{TRIP3}), (\ref{TRIP2}) becomes
\[\Di=\left(\begin{array}{cc}
\Di_{(s)}\otimes\Id_a & \gamma_3\otimes M^+\\
\gamma_3\otimes M & \Di_{(s)}\otimes \Id_b
\end{array}\right)\;.\]
Its commutator of with $\pi(f)$ is :
\be\label{TRIP6}
\bigl[\Di,\pi(f)\bigr]=
\left(\begin{array}{cc}
-\im\;\clif({\rm d}f_a)\otimes\Id_a & (f_b-f_a)\gamma_3\otimes M^+ \\
(f_a-f_b)\gamma_3\otimes M   & -\im\;\clif({\rm d}f_b)\otimes\Id_b 
\end{array}\right)\;,\ee
where the de Rham exterior differential is 
\({\rm d}f_a=\Bigl(\ev_k(f_a)\Bigr)\theta^k\) and where $\clif(\sigma^{(k)})$ denotes 
the Clifford representation of the k-form $\sigma^{(k)}$ :
\[\clif(\sigma_{i_1...i_k}\;\theta^{i_1}\wedge...\wedge\theta^{i_k})=
\sigma_{i_1...i_k}\;\gamma^{i_1}...\gamma^{i_k}\;.\]
The representation $\pi$ of (\ref{TRIP5}) extends to a $\star$-representation of 
$\Omega^\bullet(\Al)$ by :
\[\pi(f_0\dif f_1\cdots\dif f_k)=
\pi(f_0)[\Dirac,\pi(f_1)]\cdots[\Dirac,\pi(f_k)]\;.\]
From (\ref{TRIP5}) and (\ref{TRIP6}) it follows that the element 
$f\dif g\in \Omega^{(1)}(\Al)$ is represented by 
\be\label{TRIP7}
\pi(f\dif g)=
\left(\begin{array}{cc}
f_a-\im\;\clif({\rm d}g_a)\otimes\Id_a 
& f_a(g_b-g_a)\gamma_3\otimes M^+ \\
f_b(g_a-g_b)\gamma_3\otimes M & 
f_b-\im\;\clif({\rm d}g_b)\otimes\Id_b
\end{array}\right)\;,\ee 
A general element $F\in \Omega^{(1)}(\Al)$, given by $F_{\alpha\;\beta}(x,y)$, is then 
represented as an operator on $\Hi$ by :
\be\label{TRIP8}
\pi(F)=
\left(\begin{array}{cc}
-\im\;\clif(\sigma^{(1)}_a)\otimes\Id_a & 
\sigma^{(0)}_{ab}\;\gamma_3\otimes M^+ \\
\sigma^{(0)}_{ba}\;\gamma_3\otimes M & 
-\im\;\clif(\sigma^{(1)}_b)\otimes\Id_b
\end{array}\right)\;,\ee 
where the $\sigma^{(k)}$'s are differential k-forms given by :
\be\label{one-form}
\sigma_a^{(1)}(x)=\Bigl(\ev_{k,y}\,F_{aa}(x,y)\Bigr)_{\mid y=x}\;\theta^k_x\;,\;
\sigma_b^{(1)}(x)=\Bigl(\ev_{k,y}\,F_{bb}(x,y)\Bigr)_{\mid y=x}\;\theta^k_x\,,\]
\[\sigma_{ab}^{(0)}(x)=F_{ab}(x,y)_{\mid y=x}\;,\;
\sigma_{ba}^{(0)}(x)=F_{ba}(x,y)_{\mid y=x}\;.\ee
The representative of a universal 2-form $f\dif g\dif h$ will be given by the product of 
the matrix (\ref{TRIP7}) with
\[\left(\begin{array}{cc} -\im\;\clif({\rm d}h_a)\otimes\Id_a 
& (h_b-h_a)\gamma_3\otimes M^+ \\ (h_a-h_b)\gamma_3\otimes M & 
-\im\;\clif({\rm d}h_b)\otimes\Id_b \end{array}\right)\;.\]
The result is :
\[\pi(f\dif g\dif h)=\left(\begin{array}{cc}
\pi(f\dif g\dif h)_{[aa]} & \pi(f\dif g\dif h)_{[ab]}\\
\pi(f\dif g\dif h)_{[ba]} & \pi(f\dif g\dif h)_{[bb]}\end{array}\right)\;,\]
where 
\[\pi(f\dif g\dif h)_{[aa]}=
-f_a\;\clif({\rm d}g_a)\;\clif({\rm d}h_a)\otimes\Id_a
+f_a(g_b-g_a)(h_a-h_b)\otimes M^+M \;,\]
\[\pi(f\dif g\dif h)_{[ab]}=
-\im\;\Bigl(f_a\;\clif({\rm d}g_a)(h_b-h_a)
- f_a(g_b-g_a)\;\clif({\rm d}h_b)\Bigr)\gamma_3\otimes M^+ \;,\]
\[\pi(f\dif g\dif h)_{[ba]}=
-\im\;\Bigl(-f_b(g_a-g_b)\;\clif({\rm d}h_a)
+f_b\;\clif({\rm d}g_b)(h_a-h_b)\Bigr)\gamma_3\otimes M\;,\]
\[\pi(f\dif g\dif h)_{[bb]}= 
-f_b\;\clif({\rm d}g_b)\;\clif({\rm d}h_b)\otimes\Id_b
+f_b(g_a-g_b)(h_b-h_a)\otimes MM^+\;.\]
A generic universal two-form $G$ is represented by 
\be\label{TRIP9}
\pi(G)=\left(\begin{array}{cc}
\pi(G)_{[aa]} & \pi(G)_{[ab]}\\
\pi(G)_{[ba]} & \pi(G)_{[bb]}\end{array}\right)\;,\ee
with\footnote{Here we have used $\clif(\theta^k)\clif(\theta^\ell)=
\clif(\theta^k\wedge\theta^\ell) + \delta^{k\ell}$.}
\beann
\pi(G)_{[aa]}&=&-\clif(\rho_{aaa}^{(2+0)})\otimes\Id_a+
\rho_{aba}^{(0)}\otimes M^+M \;,\\
\pi(G)_{[ab]}&=&-\im\;\clif(\rho_{ab}^{(1)})\gamma_3\otimes M^+\;,\\
\pi(G)_{[ba]}&=&-\im\;\clif(\rho_{ba}^{(1)})\gamma_3\otimes M\;,\\ 
\pi(G)_{[bb]}&=&-\clif(\rho_{bbb}^{(2+0)})\otimes\Id_b+
\rho_{bab}^{(0)}\otimes MM^+\;,
\eeann
where the differential forms $\rho^{(k)}(x)$ are given by :
\bea
\rho_{aaa}^{(2+0)}(x)&=&\rho_{aaa}^{(2)}+\rho_{aaa}^{(0)}\;,\nonumber\\
\rho_{bbb}^{(2+0)}(x)&=&\rho_{bbb}^{(2)}+\rho_{bbb}^{(0)}\;,\nonumber\\
\rho_{aaa}^{(2)}(x)&=&
\Bigl({1\over 2}[\ev_{k,y}\ev_{\ell,z}-\ev_{\ell,y}\ev_{k,z}]G_{aaa}(x,y,z)\Bigr)_{\mid y=x,z=x}\,
\theta_x^k\wedge\theta_x^\ell\;,\nonumber\\
\rho_{bbb}^{(2)}(x)&=&
\Bigl({1\over 2}[\ev_{k,y}\ev_{\ell,z}-\ev_{\ell,y}\ev_{k,z}]G_{bbb}(x,y,z)\Bigr)_{\mid y=x,z=x}\,
\theta_x^k\wedge\theta_x^\ell\;,\nonumber\\
\rho_{aaa}^{(0)}(x)&=&\delta^{kl}\,
\Bigl(\ev_{k,y}\ev_{\ell,z}G_{aaa}(x,y,z)\Bigr)_{\mid y=x,z=x}\;,\nonumber\\
\tilde\rho_{bbb}^{(0)}(x)&=&\delta^{kl}\,
\Bigl(\ev_{k,y}\ev_{\ell,z}G_{bbb}(x,y,z)\Bigr)_{\mid y=x,z=x}\;,\nonumber\\
\rho_{aba}^{(0)}(x)&=&G_{aba}(x,y,z)_{\mid y=x,z=x}\;,\nonumber\\
\rho_{bab}^{(0)}(x)&=&G_{bab}(x,y,z)_{\mid y=x,z=x}\;,\nonumber\\
\rho_{ab}^{(1)}(x)&=&
\Bigl(\ev_{k,y}G_{aab}(x,y,z)-
\ev_{k,z}G_{abb}(x,y,z)\Bigr)_{\mid y=x,z=x}\,\theta_x^k\;,\nonumber\\ 
\rho_{ba}^{(1)}(x)&=&
\Bigl(\ev_{k,y}G_{bba}(x,y,z)-
\ev_{k,z}G_{baa}(x,y,z)\Bigr)_{\mid y=x,z=x}\,\theta_x^k\;.\label{two-form}
\eea
The representation $\pi$ of $\Omega^\bullet(\Al)$ is a $\star$-representation but it is not a
differential representation of $\Omega^\bullet(\Al)$ in the sense that 
$G\in Ker(\pi)={\cal J}_0$ does not imply $\pi(\dif G)=0$.
To obtain a graded differential algebra of operators in $\Hi$, it is necessary to take the quotient
of $\Omega^\bullet(\Al)$ by the graded differential ideal ${\cal J}={\cal J}_0+\dif{\cal J}_0$ 
with canonical projection 
\be\label{TRIP10}
\pi_D:\Omega^\bullet(\Al)\mapsto \Omega^\bullet_D(\Al)={\Omega^\bullet(\Al)\over{\cal J}}=
\bigoplus_{k=0}^\infty\;\Omega_D^{(k)}(\Al)\;.\ee 
Using the homomorphism theorem for algebras, it is easily seen that 
\[\Omega_D^{(k)}(\Al)={\pi\Bigl(\Omega^{(k)}(\Al)\Bigr)\over\pi(\dif{\cal J}_0^{(k-1)})}\;.\]
Since $\pi$ is a faithful representation of $\Al$, its kernel ${\cal J}_0^{(0)}$ has to be zero and 
\(\Omega^{(1)}_D(\Al)\cong\pi\Bigl(\Omega^{(1)}(\Al)\Bigr)\). So $F\in\Omega^{(1)}(\Al)$ 
has in $\Omega_D^{(1)}(\Al)$ the same representative as given by (\ref{TRIP8}).\\
To compute $\Omega^{(2)}_D(\Al)$, we need $\dif{\cal J}_0^{(1)}$. According to (\ref{TRIP8}), 
${\cal J}^{(1)}_0$ is given by :  
\[\Bigl\{F\in \Omega^{(1)}(\Al)\Vert\;F_{ab}(x,x)=0=F_{ba}(x,x)\,;\,
{\ev_{k,y}\,F_{\alpha\alpha}(x,y)}_{\mid y=x}=0,\alpha=a,b\Bigr\}\]
so that (\ref{TRIP9}) with 
$G=\dif F$ and $F\in {\cal J}_0^{(1)}$, yield a \(\pi(\dif F)\) of the form  
\[\left(\begin{array}{cc}j_a^{(0)}\otimes\Id_a & 0 \\
0 & j_b^{(0)}\otimes\Id_b\end{array}\right)\;,\] 
where \(j_\alpha^{(0)}(x)=
-\delta^{k\ell}\Bigl(\ev_{k,y}\ev_{\ell,z}F_{\alpha\alpha}(y,z)\Bigr)_{\mid y=x,z=x}\), 
with \(\alpha=a,b\). \\
In the space \(\pi\Bigl(\Omega^{(k)}(\Al)\Bigr)\), whose elements are bounded operators in
$\Hi$, the scalar product of $\pi(G_1)$ and  $\pi(G_2)$ is defined by 
\[\langle \pi(G_1);\pi(G_2)\rangle_k =
{\bf Tr}_{Dix}\Bigl\{{\pi(G_1)}^+\pi(G_2)\mid\Dirac\mid^{-d}\Bigr\}\;,\]
where ${\bf Tr}_{Dix}$ is the Dixmier trace and $d$ (here $d=2$) is the dimension 
of the spectral triple as defined in Connes' book \cite{Con1} .\\
With respect to this scalar product, \(\pi\Bigl(\Omega^{(k)}(\Al)\Bigr)\) can be completed to a
Hilbert space $\Hi^{(k)}$ and its quotient by \(\pi(\dif{\cal J}_0^{(k-1)})\), i.e.
\(\Omega^{(k)}_D(\Al)\), will be a dense subspace of  \(\pi(\dif{\cal J}_0^{(k-1)})^\perp\),
the orthogonal complement of $\pi\Bigl(\dif{\cal J}_0^{(k-1)}\Bigr)$.\\ 
Let ${\cal P}^{(k)}$ be the projector on \(\Hi^{(k)}_D=\pi(\dif{\cal J}_0^{(k-1)})^\perp\),
then a scalar product in \(\Omega^{(k)}_D(\Al)\) is defined by :
\be\label{TRIP11}
\langle \pi_D(G_1);\pi_D(G_2)\rangle_{k,D}=
\langle {\cal P}^{(k)}(\pi(G_1));{\cal P}^{(k)}(\pi(G_2))\rangle_k\;.\ee
It can also be shown that \(\Hi^{(k)}_D\) is a Hilbert $\Al$-bimodule with a two-sided
representation of the unitaries \(\;{\cal U}(\Al)=\{u\in \Al\vert uu^+=u^+u=1\}\). 
Indeed, if $G\in\Omega^{(k)}(\Al)$ then $uG$ and $Gu$ also belong to $\Omega^{(k)}(\Al)$ so that
$\Hi^{(k)}$ is a Hilbert $\Al$-bimodule. Furthermore $\pi(u)\pi_D(G)=\pi_D(uG)$ 
and $\pi_D(G)\pi(u)=\pi_D(G u)$ show that ${\cal P}^{(k)}:\Hi^{(k)}\rightarrow\Hi^{(k)}_D$ 
is a bimodule homomorphism and the Dixmier trace properties guarantee that 
\bea\label{TRIP12}
\langle \pi_D(G_1);\pi_D(G_2)\rangle_{k,D}&=&
\langle\pi(u)\pi_D(G_1);\pi(u)\pi_D(G_2)\rangle_{k,D}\nonumber\\
&=&\langle \pi_D(G_1)\pi(u);\pi_D(G_2)\pi(u)\rangle_{k,D}\;.\eea
The following trace theorems of Connes \cite{Con1} will be needed in the sequel of the calculations.

{\bf 1)} If $A_s\otimes B$ is a bounded operator in $\Hi=\Hi_{(s)}\otimes\Co^N$, then 
\be\label{TRIP13}
{\bf Tr}_{Dix}\biggl\{(A_s\otimes B)\mid\Dirac\mid^{-2}\biggr\}=
{\bf Tr}_{Dix}\Bigl\{A_s\mid\Di_{(s)}\mid^{-2}\Bigr\}\;{\bf tr}\{B\}\;,\ee
where {\bf tr} is the ordinary trace on $N\times N$ matrices.

{\bf 2)} Let $A_s$ be a section of the Clifford bundle over a compact d-dimensional 
manifold $M\;(=S^2)$ with its action on the (Pensov) spinors, then
\be\label{TRIP14}
{\bf Tr}_{Dix}\Bigl\{A_s\mid\Di_{(s)}\mid^{-d}\Bigr\}=
({1\over 4\pi})^{d/2}{1\over \Gamma(d/2+1)}\int_M\,{\bf tr}_{\clif}\{A_s\}\omega\;,\ee
where ${\bf tr}_{\clif}$ is the trace on the representation of the Clifford algebra.\\
\vspace{0.3cm}

\noindent
An element $\pi(G)$ of $\Omega_D^{(2)}(\Al)\cong \Bigl(\pi(\dif{\cal J}_0^{(1)})\Bigr)^\perp$
is of the form given in (\ref{TRIP9}) and has to obey : 
\[\forall\;j_\alpha^{(0)}\,:{\bf Tr}_{Dix}\biggl\{\left(\begin{array}{cc}
j_a^{(0)}(x)^*\otimes\Id_a & 0 \\0 & j_b^{(0)}(x)^*\otimes\Id_b\end{array}\right)
\pi(G)\mid\Dirac\mid^{-2}\biggl\}=0\;.\]
Using the trace properties (\ref{TRIP13}) and (\ref{TRIP14}), we obtain the orthogonality condition :
\be\label{Nojunk}
\rho_{aaa}^{(0)}-{1\over N_a}\rho_{aba}^{(0)}\;{\bf tr}\{M^+M\}=0\quad;\quad
\rho_{bbb}^{(0)}-{1\over N_b}\rho_{bab}^{(0)}\;{\bf tr}\{MM^+\}=0\;.\ee
Subtracting these equalities from the diagonals of (\ref{TRIP9})
yields the following representative of $\pi_D(G)\in\Omega_D^2(\Al)$ :
%%%%
%
\be\label{TRIP15}
\pi_D(G)=\left(\begin{array}{cc}
\pi_D(G)_{[aa]} & \pi_D(G)_{[ab]}\\
\pi_D(G)_{[ba]} & \pi_D(G)_{[bb]}\end{array}\right)\;,\ee
where 
\beann
\pi_D(G)_{[aa]}&=&-\clif(\rho_{aaa}^{(2)})\otimes\Id_a
+\rho_{aba}^{(0)}\otimes\Bigl[M^+M\Bigr]_{NT} \;,\\
\pi_D(G)_{[ab]}&=&-\im\;\clif(\rho_{ab}^{(1)})\gamma_3\otimes M^+\;,\\
\pi_D(G)_{[ba]}&=&-\im\;\clif(\rho_{ba}^{(1)})\gamma_3\otimes M\;,\\ 
\pi_D(G)_{[bb]}&=&-\clif(\rho_{bbb}^{(2)})\otimes\Id_b
+\rho_{bab}^{(0)}\otimes\Bigl[MM^+\Bigr]_{NT}\;.\eeann
with the traceless matrices\footnote{Note that when $N_a=N_b=N$ and $M$ is a scalar matrix, 
these traceless matrices $\Bigl[M^+M\Bigr]_{NT}$ and $\Bigl[MM^+\Bigr]_{NT}$ vanish and there is no
$\rho_\alpha^{(0)}$ term in $\pi_D(G)$. Physically this implies that, in order to have a Higgs
mechanism, a nontrivial mass spectrum is necessary!} :
\beann
\Bigl[M^+M\Bigr]_{NT}&=&M^+M-{1\over N_a}{\bf tr}\{M^+M\}\;,\\
\Bigl[MM^+\Bigr]_{NT}&=&MM^+-{1\over N_b}{\bf tr}\{MM^+\}\;.\eeann
The scalar product (\ref{TRIP11}) in $\Omega_D^2(\Al)$ is calculated using, 
besides the trace theorems, the identities: 
\beann{1\over 2^{d/2}}{\bf tr}_{\clif} \Bigl\{(\clif(\rho^{(k)}))^+\clif(\rho^{\prime(k)}\Bigr\}
&=&k!\;(\rho^{(k)}_{i_1\dots i_k})^*\rho^{\prime(k)}_{j_1\dots j_k}\;
\delta^{i_1j_1}\dots\delta^{i_kj_k}\\
&=&g^{-1}(\rho^{(k)*};\rho^{\prime(k)}).\eeann
In terms of the Hodge dual $\star$, defined by 
\[g^{-1}(\rho^{(k)*};\rho^{\prime(k)})\omega
=(\rho^{(k)})^*\wedge\star\rho^{\prime(k)}\;,\]
the scalar product reads :
\bea\label{TRIP16}
&&\langle \pi_D(G);\pi_D(G^\prime)\rangle_{2,D}=\nonumber\\
&&{1\over 2\pi}\Biggl\{
N_a\int_{S^2}{\rho^{(2)}_{aaa}}^*\wedge\star\rho^{\prime(2)}_{aaa}
+N_b\int_{S^2}{\rho^{(2)}_{bbb}}^*\wedge\star\rho^{\prime(2)}_{bbb}\nonumber\\
&& 
+{\bf tr}\{MM^+\}\;
\int_{S^2}{\rho^{(1)}_{ab}}^*\wedge\star\rho^{\prime(1)}_{ab}
+{\bf tr}\{M^+M\}\;
\int_{S^2}{\rho^{(1)}_{ba}}^*\wedge\star\rho^{\prime(1)}_{ba}\nonumber\\
&&
+{\bf tr}\Bigl\{[M^+M]_{NT}^2\Bigr\}\;
\int_{S^2}{\rho^{(0)}_{aba}}^*\wedge\star\rho^{\prime(0)}_{aba}
+{\bf tr}\Bigl\{[MM^+]_{NT}^2\Bigr\}\;
\int_{S^2}{\rho^{(0)}_{bab}}^*\wedge\star\rho^{\prime(0)}_{bab}\Biggr\}\;.\nonumber\\
&&
\eea
%
%%%%%%%%%%%%
%\newpage
%%%%%%%%%%%%
%
\subsection{The Yang-Mills-Higgs action}\label{Yang-Mills-Higgs}
The universal connection in $\Mod$, given by the matrices $\fatl A_{\alpha\beta}(x,y)\fatr$ 
of (\ref{PMod6}), is represented in $\Omega^{(1)}_D(\Al)$ by an operator of the form (\ref{TRIP8})
where the differential forms $\sigma^{(k)}_{\cdots}$ are matrix-valued.
\bea\label{YMH1}
\BL\sigma_a^{(1)}(x)\BR&=&
\BL\alpha_a(x)\BR\equiv
\Bigl(\ev_{k,y}\fatl\omega_{aa}(x,y)\fatr\Bigr)_{\mid y=x}\;\theta_x^k\;,\nonumber\\
\BL\sigma_b^{(1)}(x)\BR&=&
\alpha_b(x)\fatl P_b(x)\fatr\equiv
\Bigl(\ev_{k,y}\omega_b(x,y)\Bigr)_{\mid y=x}\;\theta_x^k\;\fatl P_b(x)\fatr\;,\nonumber\\ 
\BL\sigma_{ab}^{(0)}(x)\BR&=&
\BL\vert\Phi_{ab}(x,x)\rangle\langle\nu(x)\vert\BR\;,\nonumber\\ 
\BL\sigma_{ba}^{(0)}(x)\BR&=&
\BL\vert\nu(x)\rangle\langle\Phi_{ba}(x,x)\vert\BR\;.
\eea
The monopole connection (\ref{PMod13}) also implements a differential one-form :
\bea\label{YMH2}
\mu_b(x)&=&(\ev_{k,y}m_b(x,y))_{\mid y=x}\;\theta^k_x
=\langle\nu(x)\vert\Bigl({\rm d}\vert\nu(x)\rangle\Bigr)\;,\nonumber\\
&=&{1/2\over 1+\vert\nu(x)\vert^2}\Bigl(\nu(x)^*{\rm d}\nu(x)-\nu(x){\rm d}\nu(x)^*\Bigr)\;.
\eea
It is also convenient to introduce the Higgs field doublets :
\bea\label{YMH3}
\vert\eta_{ab}(x)\rangle&=\vert H_{ab}(x,x)\rangle&
=\vert\Phi_{ab}(x,x)\rangle+\vert\nu(x)\rangle\;,\nonumber\\
\langle\eta_{ba}(x)\vert&=\langle H_{ab}(x,x)\vert&
=\langle\Phi_{ba}(x,x)\vert+\langle\nu(x)\vert\;.\eea
The hermiticity of the connection (\ref{PMod9}) yields :
\be\label{YMH4}\BL\alpha_a\BR^+=-\BL\alpha_a\BR,\;(\alpha_b)^*=-\alpha_b,\;
(\mu_b)^*=-\mu_b,\;\langle\eta_{ba}\vert=\vert\eta_{ab}\rangle^+\;.\ee
From (\ref{PMod17}) it follows that, under an active gauge transformation, the differential forms (\ref{YMH1}) and (\ref{YMH2}) behave as :
\beann
\fatl\alpha_a^\U\fatr&=&\fatl\U_a\fatr^{-1}\fatl\alpha_a\fatr\fatl\U_a\fatr+
\fatl\U_a\fatr^{-1}{\rm d}\fatl\U_a\fatr\\
\alpha^{\U}_b&=&(\bu_b)^{-1}\alpha_b(\bu_b)+(\bu_b)^{-1}{\rm d}\bu_b=
\alpha_b+(\bu_b)^{-1}{\rm d}\bu_b\\
\mu^{\U}_b&=&(\bu_b)^{-1}\mu_b(\bu_b)=\mu_b\\
\vert\eta^{\U}_{ab}\rangle
&=&\fatl\U_a\fatr^{-1}\vert\eta_{ab}\rangle \bu_b\;,
\;\langle\eta^{\U}_{ba}\vert
=(\bu_b)^{-1}\langle\eta_{ba}\vert\fatl\U_a\fatr\;.\eeann
On the other hand, under a passive gauge transformation $H_B\rightarrow H_A$, according 
to (\ref{PMod10}) and (\ref{PMod14}), they transform as :
\beann
&&\fatl\alpha_a^A\fatr=\fatl\alpha_a^B\fatr\;,\;\alpha_b^A=\alpha_b^B\;,\\
&&\vert\eta^A_{ab}\rangle=\vert\eta^B_{ab}\rangle(c_{AB})^{-n/2}\;,\;
\langle\eta^A_{ba}\vert=(c_{AB})^{+n/2}\langle\eta^B_{ba}\vert\;,\\
&&\mu^A_b=\mu^B_b+(c_{AB})^{+n/2}{\rm d}(c_{AB})^{-n/2}=
\mu^B_b-(n/2)(c_{AB})^{-1}{\rm d}c_{AB}\;.\eeann
This means that the Higgs fields \(\{\vert\eta_{ab}\rangle\,;\,\langle\eta_{ba}\vert\}\) are
actually Pensov scalars of weight $\{- n/2\,;\,+n/2\}$ and that the monopole potential cannot be
represented by a globally defined one-form on the sphere, but adquires the inhomogeneous term
\(\;-(n/2)(c_{AB})^{-1}{\rm d}c_{AB}\) in $H_B\cap H_A$.\\
The canonical projection (\ref{TRIP10}) induces a \(\Omega^{(1)}_D(\Al)\)-valued connection 
in $\Mod$ :
\[\nabla_D:
\Mod\mapsto\Mod\otimes_\Al\Omega^{(1)}_D(\Al):X\mapsto\nabla_D X\;,\]
defined by \(\nabla_D=\Bigl(\Id_\Mod\otimes\pi_D\Bigr)\circ\nabla\).\\
From (\ref{PMod11}) and (\ref{YMH1}) it follows that :
\beann
\vert(\nabla_DX)_{aa}\rangle &=&
-\im\;\clif\biggl({\rm d}\vert f_a\rangle+\fatl\alpha_a\fatr \vert f_a\rangle\biggr)\;,\\
\vert(\nabla_DX)_{ab}\rangle &=&
\biggl(\vert\eta_{ab}\rangle f_b - \vert f_a\rangle\biggr)\gamma_3 M^+\;,\\
\vert(\nabla_DX)_{ba}\rangle &=&\vert \nu\rangle
\biggl(\langle\eta_{ba}\vert f_a\rangle-f_b\biggr)\gamma_3 M \;,\\
\vert(\nabla_DX)_{bb}\rangle &=&\vert\nu\rangle 
-\im\;\clif\biggl({\rm d}f_b +(\alpha_b+\mu_b)f_b\biggr)\;.\eeann
The curvature (\ref{PMod15}) is represented in $\Omega^{(2)}_D(\Al)$ by $\pi_D(R)$ of the 
form (\ref{TRIP15}), where the differential forms $\rho^{(k)}_{\dots}$'s are now $2\times 2$-matrix
valued.\\ The diagonal elements of $\pi_D(R)$ are given by :
\beann
\BL\rho_{aaa}^{(2)}\BR&=
&\fatl F_a\fatr={\rm d}\fatl\alpha_a\fatr+ \fatl\alpha_a\fatr\wedge\fatl\alpha_a\fatr\;,\nonumber\\
\BL\rho_{bbb}^{(2)}\BR&=
&F_b\fatl P_b\fatr=({\rm d}\alpha_b+{\rm d}\mu_b)\fatl P_b\fatr\;,\nonumber\\
\BL\rho_{aba}^{(0)}\BR&=
&\vert\eta_{ab}\rangle\langle\eta_{ba}\vert - \fatl{\bf Id}\fatr\;,\nonumber\\
\BL\rho_{bab}^{(0)}\BR&=
&(\langle\eta_{ba}\vert\eta_{ab}\rangle - 1)\;\fatl P_b\fatr\;.\eeann
The off-diagonal elements are given in terms of the covariant differentials 
of the Higgs fields (\ref{YMH3}):
\bea\label{YMH5}
\vert\nabla\eta_{ab}\rangle &=&
{\rm d}\;\vert\eta_{ab}\rangle +\fatl\alpha_a\fatr\vert \eta_{ab}\rangle
- \vert \eta_{ab}\rangle(\alpha_b+\mu_b)\;,\nonumber\\
\langle\nabla\eta_{ba}\vert &=&
{\rm d}\;\langle \eta_{ba}\vert -\langle \eta_{ba}\vert \fatl\alpha_a\fatr
+(\alpha_b+\mu_b)\langle \eta_{ba}\vert\;.\eea
They read
\[
\BL\rho_{ab}^{(1)}\BR=\vert \nabla\eta_{ab}\rangle\langle\nu\vert\;,\;
\BL\rho_{ba}^{(1)}\BR=\vert\nu\rangle\langle \nabla\eta_{ba}\vert\;.\]
The Yang-Mills-Higgs action is constructed as :
\bea\label{YMH6}
{\bf S}_{YMH}(\nabla_D)&=&
\lambda\,{\bf tr}_{matrix}\Bigl\{\langle\pi_D(R);\pi_D(R)\rangle_{2,D}\Bigr\}\nonumber\\
&=&\lambda\,{\bf tr}_{matrix}
\biggl\{{\bf Tr}_{Dix}\Bigl\{\pi_D(R)^{\dagger}\pi_D(R)\vert\Dirac\vert^{-2}\Bigr\}\biggr\}
\;,\eea
where $\lambda$ is a coupling constant and ${\bf tr}_{matrix}$ is the 
trace of the $2\times 2$ matrices, product of matrices $\BL\rho^{(k)}\BR^+$ with $\BL\rho^{(k)}\BR$. Since the curvature transforms as $ R\rightarrow R^\U=\U^{-1}R\U$, the gauge invariance of the action follows from the obvious extension of the representation (\ref{TRIP12}) of the unitaries in $\Hi^{(2)}_D$.
With the scalar product given by (\ref{TRIP16}), the action (\ref{YMH5}) reads :
\bea\label{YMH7}
{\bf S}_{YMH}(\nabla_D)&=&{\lambda\over 2\pi}\biggl\{
N_a\int_{S^2}{\bf tr}_{matrix}\Bigl\{\fatl F_a\fatr^+\wedge\star\fatl F_a\fatr\Bigr\}\nonumber\\&&
\quad\quad+N_b\int_{S^2}(F_b)^*\wedge\star F_b\nonumber\\&&
\quad\quad+2\;{\bf tr}\{MM^+\}\;
\int_{S^2}\langle\nabla\eta_{ba}\vert\wedge\star\vert\nabla\eta_{ab}\rangle\nonumber\\&&
\quad\quad+{\bf tr}\Bigl\{{\Bigl[M^+M\Bigr]_{NT}}^2\Bigr\}\;
\int_{S^2}\star\Bigl((\langle \eta_{ba}\vert \eta_{ab}\rangle-1)^2+1\Bigr)\nonumber\\&&
\quad\quad+{\bf tr}\Bigl\{{\Bigl[MM^+\Bigr]_{NT}}^2\Bigr\}\;
\int_{S^2}\star(\langle \eta_{ba}\vert \eta_{ab}\rangle-1)^2
\biggr\}\;.\eea
%
%
%%%%%%%%%%%%
%%%\newpage
%%%%%%%%%%%%
%
%
\subsection{The Hilbert space of particle states and the covariant Dirac operator}\label{Matter}
The tensor product over $\Al$ of the right $\Al$-module $\Mod$ with the (left-module) Hilbert space
$\Hi$ is itself a Hilbert space $\Hi_{\bf p}=\Mod\otimes_\Al\Hi$, with scalar product induced by the scalar product 
(\ref{TRIP4}) in $\Hi$ and the hermitian structure ${\bf h}$ in the module $\Mod$ :
\[(X\otimes_\Al \BPsi;Y\otimes_\Al\BPhi) = (\BPsi;\pi({\bf h}(X,Y))\BPhi)\;.\]
A generic element of $\Hi_{\bf p}$ can be written as \(\Vert\BPsi_{\bf p}\BRR=E_i\otimes_\Al\BPsi^i\), where $\BPsi^i\in\Hi$ obeys $\pi({P^i}_j)\BPsi^j=\BPsi^i$.
In the model considered here, $\Hi=\Hi_{(s)}\otimes\Bigl(\Co^{N_a}\oplus\Co^{N_b}\Bigr)$ and the projective module is $\Mod=\Proj \Al^2$, with $\Proj$ defined by the homotopy class $[g]$ in (\ref{PMod3}). A state \(\Vert\BPsi_{\bf p}\BRR\) describing particles, is thus represented by :
\begin{enumerate}
\item A pair of Pensov spinors of $\Hi_{(s)}$, given by :\\ 
\(\vert\psi_a(x)\rangle=\left(\begin{array}{c}\psi^{1}_a(x)\\ \psi^{2}_a(x)\end{array}\right)\), each with $N_a$ values of the generation index.
\item A single Pensov spinor $\psi_b(x)$ of $\Hi_{(s+n/2)}$, with a $N_b$-valued generation index, such that 
\(\vert\psi_b(x)\rangle=\left(\begin{array}{c}\psi^{1}_b(x)\\ \psi^{2}_b(x)\end{array}\right)=\vert\nu(x)\rangle\;\psi_b(x)$ in $H_B$.
\end{enumerate}
The $\star$-representation $\pi$ of $\Omega^\bullet(\Al)$ in $\Hi$ induces a mapping 
\be\label{Pi1}
\pi_1:\Mod\otimes_\Al\Omega^\bullet(\Al)\mapsto {\cal B}(\Hi,\Hi_{\bf p}):
X\otimes_\Al F\mapsto\pi_1(X\otimes_\Al F)\;,\ee
where \({\cal B}(\Hi,\Hi_{\bf p})\) are the bounded linear operators from $\Hi$ to $\Hi_{\bf p}$.\\
It is defined by \(\pi_1(X\otimes_\Al F)\BPsi=X\otimes_\Al\pi(F)\BPsi\;.\)\\
Furthermore, there is a mapping 
\be\label{Pi2}
\pi_2:{\rm HOM}_\Al(\Mod,\Mod\otimes_\Al\Omega^\bullet)\mapsto{\cal B}(\Hi_{\bf p}):{\bf T}\mapsto\pi_2({\bf T})\;,\ee
defined by \(\pi_2({\bf T})\Bigl(X\otimes_\Al\BPsi\Bigr)=\pi_1({\bf T}X)\BPsi\).\\ 
The covariant Dirac operator in $\Hi_{\bf p}$ is defined, using (\ref{Pi1}), as 
\be\label{CovDirac1}
\Dirac_\nabla\Bigl(X\otimes_\Al\BPsi\Bigr)=X\otimes_\Al\Dirac\BPsi+\pi_1(\nabla X)\BPsi\;.\ee
It is easy to check that \(\Dirac_\nabla\Bigl(X\,f\otimes_\Al\BPsi\Bigr)=
\Dirac_\nabla\Bigl(X\otimes_\Al\pi(f)\BPsi\Bigr)\) so that $\Dirac_\nabla$ is well defined in $\Hi_{\bf p}$.\\
A grading in $\Hi_{\bf p}$ is defined by 
\be\label{partgrad}
\Gamma_{\bf p}: X\otimes_\Al\BPsi\mapsto X\otimes_\Al \Gamma\BPsi\;\ee
and the covariant Dirac operator is odd with respect to this grading :
\be\label{oddness}
\Dirac_\nabla\Gamma_{\bf p}+\Gamma_{\bf p}\Dirac_\nabla=0
\ee
With $\Vert\BPsi_{\bf p}\BRR$ as above, \(\Dirac_\nabla\) is calculated as follows.
\[\Dirac_\nabla\Vert\BPsi_{\bf p}\BRR=
E_i\otimes_\Al\Bigl(\Di_\nabla\Vert\BPsi_{\bf p}\BRR\Bigr)^i\;,\]
 where
\(\Bigl(\Di_\nabla\Vert\BPsi_{\bf p}\BRR\Bigr)^i=
\Bigl(\pi({P^i}_j)\Dirac+\pi({\fatl A\fatr^i}_j)\Bigr)\BPsi^j\) is represented by
\be\label{CovDirac2}
\left(\begin{array}{cc}
(\Di_\nabla)_{aa}&(\Di_\nabla)_{ab}\\
(\Di_\nabla)_{ba}&(\Di_\nabla)_{bb}\end{array}\right)
\left(\begin{array}{l}\vert\psi_a\rangle\\ \vert\nu\rangle\psi_b\end{array}\right)\;,\ee
\beann
\Bigl(\Di_\nabla\Bigr)_{aa}&=&
\Di_{(s)}\otimes\Id_a-\im\;\clif\fatl\alpha_a\fatr\otimes\Id_a\;,\\
\Bigl(\Di_\nabla\Bigr)_{ab}&=&
\fatl\vert\eta_{ab}\rangle\langle\nu\vert\fatr\otimes\gamma_3 M^+\;,\\
\Bigl(\Di_\nabla\Bigr)_{ba}&=&
\fatl\vert\nu\rangle\langle\eta_{ba}\vert\fatr\otimes\gamma_3 M\;,\\
\Bigl(\Di_\nabla\Bigr)_{bb}&=&
\fatl P_b\fatr\Di_{(s)}\fatl P_b\fatr
\otimes\Id_b-\im\;\clif(\alpha_b)\fatl P_b\fatr\;.\eeann
Now, \(\langle\nu\vert\Di_{(s)}\vert\nu\rangle\psi_b=\Di_{(s)}\psi_b-\im\;\clif(m_b)\psi_b\) and 
with our choice (\ref{PMod3}) of the representative of the homotopy class $[n]\in\Ent$, 
we obtain\footnote{Note that with a different choice in (\ref{PMod3}), a globally defined 
differential one-form would be added to \(\Di_{(s+n/2)}\) and this can always be absorbed in
\(\alpha_b\).}
\be\label{SfromI}
\fatl P_b\fatr\Di_{(s)}\vert\nu\rangle\psi_b=\vert\nu\rangle\Di_{(s+n/2)}\psi_b\;.\ee 
Substituting (\ref{SfromI}) and (\ref{CovDirac2}) in (\ref{CovDirac1}) yields finally:
\bea\label{CovDirac3}
\Bigl(\Di_\nabla\Vert\BPsi_{\bf p}\BRR\Bigr)_a&=&
\Bigl(\Di_{(s)}-\im\;\clif\fatl\alpha_a\fatr\Bigr)\vert\psi_a\rangle
+\vert \eta_{ab}\rangle\gamma_3 M^+\psi_b\;,\nonumber\\
\Bigl(\Di_\nabla\Vert\BPsi_{\bf p}\BRR\Bigr)_b&=&
\vert\nu\rangle\Bigl[\Bigl(\Di_{(s+n/2)}
-\im\;\clif(\alpha_b)\Bigr)\psi_b
+\gamma_3 M\langle \eta_{ba}\vert\psi_a\rangle\Bigr]\;.\eea
The matter action functional is then constructed as ;
\bea\label{Matiere}
{\bf S}_{\bf Mat}(\Vert\BPsi_{\bf p}\BRR,\nabla_D)&=&
\Bigl(\Vert\BPsi_{\bf p}\BRR;\Dirac_\nabla\Vert\BPsi_{\bf p}\BRR\Bigr)\nonumber\\
&=&\int_{S^2}\star\biggl\{
\langle\psi_a\vert\Bigl(\Di_{(s)}-\im\;\clif\fatl\alpha_a\fatr\Bigr)
\vert\psi_a\rangle\nonumber\\
&&+\langle\psi_a\vert\eta_{ab}\rangle\gamma_3 M^+\psi_b
+(\psi_b)^+\gamma_3 M\langle \eta_{ba}\vert\psi_a\rangle\nonumber\\
&&+(\psi_b)^+\Bigl(\Di_{(s+n/2)}-\im\;\clif(\alpha_b)\Bigr)\psi_b
\biggr\}\;.\eea
The hermiticity condition (\ref{YMH4}) guarantees that the action is real.
If an Euclidean chiral theory is aimed for, then  
\(\Gamma_{\bf p}\Vert\BPsi_{\bf p}\BRR=\Vert\BPsi_{\bf p}\BRR\) implies that the action
 (\ref{Matiere}) vanishes identically due to the oddness of the Dirac operator (\ref{oddness}). 
A proposed way out, as in \cite{V-GB}, is just to make an easy switch going to an indefinite
 Minkowski type metric changing the $\psi^+$ to a $\bar\psi=\psi^+ \gamma^0$ so that the presence
of the $\gamma^0$ provides an extra factor minus one and the action does not vanish. Such a "usual
incantation " \cite{M-GB-V} appears highly unaesthetic and rather unsatisfactory. It seems
necessary to double the Hilbert space in order to deal with this issue. This can be achieved
introducing a Hilbert space $\Hi_{\bf \bar p}$ of "anti-particle" states. The need of doubling the
Fermion fields also arises in the usual Euclidean quantum field theory,
 where the fermion fields are operator valued in Fock space, in order to cure inconsistent
hermiticity properties of the propagators \cite{O-S,F-O}. Alternative proposals were made by
\cite{Met} and more recently by \cite{vN-W}. Related comments
 by \cite{L-M-M-S} in a  non-commutative geometric setting, should also be mentioned.\\
Here, however, we choose to remain with the primary interpretation of \(\Vert\BPsi_{\bf p}\BRR\) 
as a {\it state} in the Hilbert space \(\Hi_{\bf p}\) represented by Euclidean {\it wave
functions}. This means that in this work we endeavour an Euclidean one-particle (plus one would-be
anti-particle) field theory, which, in a path integral formalism, may hopefully lead to a proper
quantum theory.
%%%%%%%
%%%%%%%
\newpage
%%%%%%%
%%%%%%%
\section{Real spectral triples}\label{Real Triples}
\setcounter{equation}{0} 
\subsection{The real Pensov-Dirac spectral triple}\label{RPDtriple}
The complex conjugation $\Ko$ transforms a Pensov field of weight $s$, $\sigma^{(s)}$, into a 
Pensov field $\sigma^{(s)*}$ of weight $-s$. Besides the Hilbert space of Pensov spinors
\(\psi_{(+)}\) of weight $s$, denoted here as 
$\Hi_{1(+)}$, we also introduce $\Hi_{1(-)}$, with spinors \(\psi_{(-)}\) of weight $-s$.
A real structure will be induced by a pair of anti-linear mappings 
$\J_{1(\pm)}:\Hi_{1(\pm)}\mapsto\Hi_{1(\mp)}$, which are  required to preserve the real
Clifford-algebra module structure of the spinor spaces : 
\bea
\J_{1(\pm)}\lambda\psi_{(\pm)}&=&\bar\lambda\J_{1(\pm)}\psi_{(\pm)}\nonumber\\
\J_{1(\pm)}\left(\gamma^k\psi_{(\pm)}\right)&=
&\alpha\;\gamma^k\left(\J_{1(\pm)}\psi_{(\pm)}\right)\label{J1}\;,\eea
where we allow for $\alpha$ to be a sign factor $\pm 1$.\\
With 
\be\label{J2}
\J_{1(\pm)}\psi_{(\pm)}=a_{(\pm s)}\;\Fu_{1,\alpha}\,\Ko\;\psi_{(\pm)}\;,\ee
where $a_{(\pm s)}$ is an arbitrary complex number, we should have 
\[\Fu_{1,\alpha}\left(\gamma^k\right)^*\Fu_{1,\alpha}^{-1}=\alpha\;\gamma^k\;.\]
In the chiral representation 
\(\gamma^1=\left(\begin{array}{cc} 0 & 1\\ 1 &
0\end{array}\right)\;;\;\gamma^2=\left(\begin{array}{cc} 0 & \im\\ -\im &
0\end{array}\right)\), we may choose 
$\Fu_{1,\alpha}=\left(\begin{array}{cc} 0 & 1\\ \alpha & 0
\end{array}\right)$. \\  The adjoint of \(\J_{1(\pm)}\) is given by :
\be\label{J3}
\left(\J_{1(\pm)}\right)^+\psi_{(\mp)}=a_{(\pm s)}\;\Fu_{1,\alpha}^t\,\Ko\;\psi_{(\mp)}\;.\ee
Demanding that 
\be\label{J4}
\J_{1(\pm)}\left(\J_{1(\pm)}\right)^+=
\Id_{(\mp)}\;,\;\left(\J_{1(\pm)}\right)^+\J_{1(\pm)}=\Id_{(\pm)}\;,\ee
restricts $a_{(\pm s)}$ be a phase factor. On \(\Hi_{1}=\Hi_{1(+)}\oplus\Hi_{1(-)}\), the antilinear isometry defined by
\be\label{J5}
\J_{1}=\left(\begin{array}{cc} 0 &\J_{1(-)} \\ \J_{1(+)} & 0 \end{array}\right)\;\ee
obeys 
\[\J_{1}\left(\begin{array}{cc}
\gamma^k & 0 \\ 0 & \gamma^k\end{array}\right)= \alpha \; 
\left(\begin{array}{cc}
\gamma^k & 0 \\ 0 & \gamma^k\end{array}\right)\J_{1}\;.\]
If we require 
\be\label{J6}
{\J_{1}}^2=\epsilon_1\;\Id_1\;,\;\hbox{with}\;\epsilon_1=\pm 1,\ee
the phases $a_{(\pm s)}$ are related by $a_{(-s)}=\alpha\;\epsilon_1\;a_{(+s)}$ and  
\(\J_{1(-)}=\epsilon_1\left(\J_{1(+)}\right)^+\).
The antilinear mappings $\J_{1(\pm)}$ intertwine with the Dirac operators $\Di_{(\pm s)}$ as :
\be\label{J7}
\J_{1(\pm)}\Di_{(\pm s)}=-\alpha\;\Di_{(\mp s)}\J_{1(\pm)}\;.
\ee
On $\Hi_{1(+)}$, the Dirac operator is chosen as $\Di_{1(+)}=\Di_{(s)}$, but on $\Hi_{1(-)}$ we may choose $\Di_{(-s)}$ up to a sign. Let $\epsilon_1^{\;\prime}$ be another arbitrary sign factor, then the choice\footnote{For notational convenience, we define $\kappa(+)=+1$ and $\kappa(-)=-\alpha\;\epsilon_1^{\;\prime}$ so that $\Di_{1(\pm)}=\kappa(\pm)\;\Di_{(\pm s)}$.} 
\be\label{J8}
\Di_{1(-)}=-\alpha\;\epsilon_1^{\;\prime}\;\Di_{(-s)}\;,\ee
yields a Dirac operator 
\(\Di_1=\left(\begin{array}{cc}\Di_{1(+)} & 0 \\ 0 & \Di_{1(-)}\end{array}\right)\)
intertwining with $\J_1$ as :
\be\label{J9}
\J_1\;\Di_1=\epsilon_1^{\;\prime}\;\Di_1\J_1\;.\ee
The representation of $\Al_1=\Fu(S^2;\Co)$ in $\Hi_{1}$ is obtained by taking two copies of the representation in $\Hi_{(s)}$ :
\be\label{J10}
\pi_1(f)\;\left(\begin{array}{c}\psi_{(+)}\\ \psi_{(-)}\end{array}\right)(x)=
\left(\begin{array}{cc} f(x) & 0 \\
0 & f(x) \end{array}\right)
\left(\begin{array}{c}\psi_{(+)}(x)\\ \psi_{(-)}(x)\end{array}\right)\;.\ee
In general, a real structure $\J_{1}$ induces a representation of the opposite algebra $\Al_1^{o}$ by :
\[\pi^{o}_{1}(f)=\J_{1}\left(\pi_1(f)\right)^+\J_{1}^+\;,\]
so that the Hilbert space $\Hi_{1}$ becomes an $\Al_1$ bimodule. Here $\Al_1$ is abelian, and with the representation  $\pi_1$ above, we have $\pi^{o}_1(f)=\pi_1(f)$.\\
Since 
\(\Bigl[D_{1},\pi_{1}(f)\Bigl]= 
\left(\begin{array}{cc} -\im\kappa_{(+)}\clif(df) & 0 \\ 0 & -\im\kappa_{(-)}\clif(df)\end{array}\right)\), the first-order condition
\be\label{J11}\Bigl[\Bigl[\Di_{1},\pi_{1}(f)\Bigr],\pi_{1}^{o}(g)\Bigr]=0\;,\ee
which is needed to define a connection in bimodules \cite{Dubois}, is satisfied.\\
Since \(\J_{1(\pm)}\gamma_3=-\gamma_3\J_{1(\pm)}\), the chirality in $\Hi_{1(+)}\oplus\Hi_{1(-)}$ will be taken as 
\be\label{J12}
\chi_{1}=\left(\begin{array}{cc}\gamma_3 & 0 \\ 0 & \gamma_3 \end{array}\right)\;.\ee
With this choice :
\be\label{J13}
\J_{1}\chi_{1}=\epsilon_1^{\;\prime\prime}\;\chi_{1}\J_{1}\;,\;\hbox{with}\;\epsilon_1^{\;\prime\prime}=-1\;.\ee
The $\epsilon$-sign table of Connes \cite{Con2,Con3}, corresponding to $n=2$, can be satisfied if we choose 
$\epsilon_1=-1$ and $\epsilon_1^{\;\prime}=+1$, but for the moment we shall leave these choices open.\\
The spectral triple \(\Kcyc_1=\left\{\Al_1,\Hi_{1},\Di_{1},\chi_{1},\J_1\right\}\) is actually a 0-sphere real spectral triple as defined in \cite{Con2}. For our pragmatic purposes, an $S^0$-real spectral triple may be defined as a real spectral triple with an hermitian involution $\sigma^0$ commuting with $\pi(\Al_1),\Di_{1},\chi_{1}$ and anticommuting with $\J_{1}$.\\
It is implemented by the decomposition $\Hi_{1}=\Hi_{1(+)}\oplus\Hi_{1(-)}$, which it is given in, by : 
\be\label{J14}
\sigma^0=\left(\begin{array}{cc} 
\Id_{1(+)} & 0\\
0 & -\Id_{1(-)}
\end{array}\right)\;.\ee
The doubling of the Hilbert space is justified if we interpret the Pensov spinors of $\Hi_{(s)}$ as usual (Euclidean!) Dirac spinors interacting with a magnetic monopole of strenght $s$. It seems then natural to consider the (Euclidean!) anti-particle fields as Dirac spinors "seeing" a monopole of strenght $-s$ i.e. as Pensov spinors of $\Hi_{(-s)}$.
\subsection{The real discrete spectral triple}\label{Rdtriple}
Proceeding further, as in section {\bf\ref{Triple}}, we have to compose the above $S^0$-real "Dirac-Pensov" spectral triple
\(\Kcyc_1\) with a real discrete spectral triple \(\Kcyc_2=\left\{\Al_2,\Hi_{2},\Di_{2},\chi_{2},\J_2\right\}\) over the algebra \(\Al_2=\Co\oplus\Co\). 
The most general finite Hilbert space allowing a \(\Al_2\)-bimodule structure\footnote{For a general discussion on real discrete spectral triples, we refer to (\cite{Kraj},\cite{Pasch})} is given by the direct sum
\be\label{J15}
\Hi_2=\bigoplus_{\alpha,\beta} \Co^{\;N_{\alpha\beta}}\;,\ee
where $\alpha$ and $\beta$ vary over \{a,b\} and where $N_{\alpha\beta}$ are integers.\\
Its elements are of the form
\[\left(\xi\right)=\left(\begin{array}{c}\xi^{aa} \\ \xi^{ab} \\ \xi^{ba} \\ \xi^{bb}\end{array}\right)\;,\]
where each $\xi^{\alpha\beta}$ is a column vector with $N_{\alpha\beta}$ rows.
An element $\lambda=(\lambda_a,\lambda_b)$ of $\Al_{1}$ acts on the left on $\Hi_{2}$ by:
\be\label{J16}
\pi_2(\lambda)=
\left(\begin{array}{cccc}
\lambda_a\;\Id_{N_{aa}} & 0 & 0 & 0\\
0 & \lambda_a\;\Id_{N_{ab}} & 0 & 0\\
0 & 0 &\lambda_b\;\Id_{N_{ba}}  & 0\\
0 & 0 & 0 & \lambda_b\;\Id_{N_{bb}}
\end{array}\right)\;,\ee
and on the right by:
\be\label{J17}
\pi_2^{o}(\lambda)=
\left(\begin{array}{cccc}
\lambda_a\;\Id_{N_{aa}} & 0 & 0 & 0\\
0 & \lambda_b\;\Id_{N_{ab}} & 0 & 0\\
0 & 0 &\lambda_a\;\Id_{N_{ba}}  & 0\\
0 & 0 & 0 & \lambda_b\;\Id_{N_{bb}}
\end{array}\right)\;.\ee
Although $\Al_{2}$ is an abelian algebra and, as such, isomorphic to its opposite algebra , it is not a simple algebra so that, in general, $\pi_2(\lambda)\neq\pi_2^{o}(\lambda)$. The discrete real structure, given by $\J_{2}=\Fu_{2}\Ko$, relates both by
$\pi^{o}(\lambda)=\J_{2}\pi(\lambda)^+\J_{2}^{-1}$ so that $\Fu_{2}$ is an intertwining operator for the two representations:
\be\label{J18}
\pi_2^{o}(\lambda)=\Fu_{2}\pi_2(\lambda)\Fu_{2}^{-1}\;.\ee
This requires that $N_{ab}=N_{ba}\doteq N$ and, since we require $\J_2$ to be anti-unitary, the basis in $\Hi_2$ may be chosen such that :
\be\label{J19}
\Fu_{2}=\left(\begin{array}{cccc}
\Id_{N_{aa}} & 0 & 0 & 0 \\
0 & 0 & \Id_{N} & 0 \\
0 & \Id_{N} & 0 & 0 \\
0 & 0 & 0 & \Id_{N_{bb}}
\end{array}\right)\;.\ee
This implies that :
\be\label{J20}
{\J_2}^2=\epsilon_2\;\Id_2\;,\;\hbox{with}\;\epsilon_2=+1\;.\ee
The chirality, $\chi_{2}$, defining the orientation of the spectral triple is the image of a Hochschild 0-cycle, i.e. an element of $\Al_{2}\otimes\Al^{o}_{2}$. This implies that $\chi_{2}$ is diagonal and $\chi_{\alpha\beta}=\pm 1$ on each subspace $\Co^{N_{\alpha\beta}}$.\\
Furthermore, demanding that\footnote{If we should require that $\epsilon_2^{\;\prime\prime}=-1$, then $N_{aa}=N_{bb}=0$ and $\chi_{ab}=-\chi_{ba}$ and the corresponding odd Dirac operator would not satisfy the first order condition.}
\be\label{J21}
\J_{2}\chi_{2}=\epsilon_2^{\;\prime\prime}\chi_{2}\J_{2}\;,\;\hbox{with}\;\epsilon_2^{\;\prime\prime}=+1\;,\ee 
requires $\chi_{ab}=\chi_{ba}=\chi^\prime$ so that the chirality in $\Hi_{2}$ reads:
\be\label{J22}
\chi_{2}=\left(\begin{array}{cccc}
\chi_{aa}\Id_{N_{aa}} & 0 & 0 & 0\\
0 & \chi^\prime\Id_{N} & 0 & 0\\
0 & 0 & \chi^\prime\Id_{N} & 0\\
0 & 0 & 0 & \chi_{bb}\Id_{N_{bb}}
\end{array}\right)\;.\ee
We consider the following three possibilities leading to a non trivial hermitian Dirac operator, odd with respect to this chirality :
\begin{description}
\item [2.a] $+\chi_{aa}=+\chi^\prime=-\chi_{bb}=\pm 1$
\item [2.b] $-\chi_{aa}=+\chi^\prime=+\chi_{bb}=\pm 1$
\item [2.c] $-\chi_{aa}=+\chi^\prime=-\chi_{bb}=\pm 1$
\end{description}
The corresponding Dirac operators have the form 
\begin{description}
\item [2.a,2.b]
\[\Di_{2.a}=\left(\begin{array}{cccc}
0 & 0 & 0 & K^+\\
0 & 0 & 0 & A^+\\
0 & 0 & 0 & B^+\\
K & A & B & 0
\end{array}\right)\;;\;
\Di_{2.b}=\left(\begin{array}{cccc}
0 & B^+ & A^+ & K^+\\
B & 0 & 0 & 0\\
A & 0 & 0 & 0\\
K & 0 & 0 & 0
\end{array}\right)\]
\item [2.c]
\[\Di_{2.c}=\left(\begin{array}{cccc}
0 & B^{\prime +} & A^{\prime +} & 0\\
B^\prime & 0 & 0 & A^+\\
A^\prime & 0 & 0 & B^+\\
0 & A & B & 0
\end{array}\right)\;.\]
\end{description}
The first-order condition $[[\Di_{2},\pi_2(\lambda)],\pi_2^{o}(\mu)]=0$, satisfied in case {\bf 2.c)}, implies that in case {\bf 2.a)} and {\bf 2.b)} $K$ must vanish.\\ 
If we asssume 
\be\label{J23}\J_{2}\Di_{2}=\epsilon_2^\prime\;\Di_{2}\J_{2}\;,
\;\hbox{with}\;\epsilon_2^\prime=+1\;,\ee
then
\be\label{J24}
B=A^*\;;\,B^\prime=A^{\prime\,*}\;.\ee
It should be stressed that, in order to have a non trivial Dirac operator, necessarily $N\neq 0$. This confirms that the discrete Hilbert space $\Hi_{dis}$ used in {\bf \ref{Triple}} does not allow for a real structure in the above sense. At last, it can be shown\cite{Kraj,Pasch} that noncommutative Poincar\'{e} duality, in the discrete case, amounts to the non degeneracy of the intersection matrix with elements $\bigcap_{\alpha\beta}=\chi_{\alpha\beta}N_{\alpha\beta}$. This non degeneracy condition in case {\bf 2.a)} and {\bf 2.b)} reads $N_{aa}N_{bb}+N^2\neq 0$ and is always satisfied. 
In case {\bf 2.c)} it is required that $N_{aa}N_{bb}-N^{2}\neq 0$ and if all $N$'s should be equal, this would not be  satisfied. If we insist on equal $N$'s, which is not strictly necessary, we are limited to the models {\bf 2.a)} and {\bf 2.b)} with representations given by (\ref{J16}), (\ref{J17}), Dirac operators and chirality assignments by : 
\be\label{J25}
\Di_{2.a}=\left(\begin{array}{cccc}
0 & 0 & 0 & 0\\
0 & 0 & 0 & A^+\\
0 & 0 & 0 & B^+\\
0 & A & B & 0
\end{array}\right)\;;\;
\chi_{2.a}=\chi^{\;\prime}\left(\begin{array}{cccc}
\Id_{N} & 0 & 0 & 0\\
0 & \Id_{N} & 0 & 0\\
0 & 0 & \Id_{N} & 0\\
0 & 0 & 0 & -\Id_{N}
\end{array}\right)\;.\ee
%%%
\be\label{J26}
\Di_{2.b}=\left(\begin{array}{cccc}
0 & B^+ & A^+ & 0\\
B & 0 & 0 & 0\\
A & 0 & 0 & 0\\
0 & 0 & 0 & 0
\end{array}\right)\;;\;
\chi_{2.b}=\chi^{\;\prime}\left(\begin{array}{cccc}
-\Id_{N} & 0 & 0 & 0\\
0 & \Id_{N} & 0 & 0\\
0 & 0 & \Id_{N} & 0\\
0 & 0 & 0 & \Id_{N}
\end{array}\right)\;.\ee
\subsection{The product}\label{Prodtriple}
The product of \(\Kcyc_1\) with \(\Kcyc_2\) yields the total triple \(\Kcyc=\left\{\Al,\Hi,\Di,\chi,\J\right\}\)
with algebra $\Al=\Al_1\otimes\Al_2$. Since $\Kcyc_1$ is $S^0$-real, the product is also $S^0$-real with hermitian involution \(\Sigma^0=\sigma^0\otimes\Id_2\). The total Hilbert space $\Hi=\Hi_1\otimes\Hi_2$ is decomposed as 
\be\label{J27}
\Hi=\Hi_{(+)}\oplus\Hi_{(-)}\;,\ee
where $\Hi_{(\pm)}=\Hi_{1(\pm)}\otimes\Hi_2$. Elements of $\Hi_{(\pm)}$ are represented by column matrices of the form 
\be\label{J28}
\Psi_{(\pm)}(x)=\left(\begin{array}{c}
\psi_{(\pm)aa}(x) \\ \psi_{(\pm)ab}(x) \\ \psi_{(\pm)ba}(x) \\ \psi_{(\pm)bb}(x) \end{array}\right)\;,\ee
where each $\psi_{(\pm)\alpha\beta}(x)$ is a Pensov spinor of $\Hi_{1(\pm)}$ with $N$ "internal" indices (not explicitely written down).\\
The total Dirac operator is 
\be\label{J29}
\Di=\Di_1\otimes\Id_2+\chi_1\otimes\Di_2\;,\ee
and the chirality is given by 
\be\label{J30}
\chi=\chi_1\otimes\chi_2\;.\ee
The continuum spectral triple $\Kcyc_1$ of {\bf \ref{RPDtriple}} is of dimension two and its real structure $\J_1$ obeys
\[{\J_1}^2=\epsilon_1\;\Id_1\;;\;\J_1\Di_1=\epsilon_1^{\;\prime}\Di_1\J_1\;;\; \J_1\chi_1=\epsilon_1^{\;\prime\prime}\chi_1\J_1\;,\]
where $\epsilon_1^{\;\prime\prime}=-1$ was fixed but $\epsilon_1$ and $\epsilon_1^{\;\prime}$ were independent free $\pm 1$ factors. \\
On the other hand the discrete triple $\Kcyc_2$ of {\bf \ref{Rdtriple}} is of zero dimension and $\J_2$ obeys 
\[{\J_2}^2=\epsilon_2\;\Id_2\;;\;\J_2\Di_2=\epsilon_2^{\;\prime}\Di_2\J_2\;;\; \J_2\chi_2=\epsilon_2^{\;\prime\prime}\chi_2\J_2\;,\] with $\epsilon_2=\epsilon_2^{\;\prime}=\epsilon_2^{\;\prime\prime}=+1$.\\
The real structure $\J$ of the product triple should obey 
\be\label{J31}
{\J}^2=\epsilon\;\Id\;;\;\J\Di=\epsilon^{\;\prime}\Di\J\;;\;\J\chi=\epsilon^{\;\prime\prime}\chi\J\;.\ee
If we require that Connes' sign table be satisfied, i.e. 
\begin{itemize}
\item for $\Kcyc_1$, $n_1=2$ and $\epsilon_1=-1\;,\;\epsilon_1^{\;\prime}=+1\;,\;\epsilon_1^{\;\prime\prime}=-1$,
\item for $\Kcyc_2$, $n_2=0$ and $\epsilon_2=\;\epsilon_2^{\;\prime}=\epsilon_2^{\;\prime\prime}=+1$ for $n_2=0$,  
\item for the product $\Kcyc$, $n=2$ and $\epsilon=-1\;,\;\epsilon^{\;\prime}=+1\;,\;\epsilon^{\;\prime\prime}=-1$,
\end{itemize}
it is seen that, with $\J=\J_1\otimes\J_2$, the sign table for the product is not obeyed, since such a $\J$ implies the consistency conditions
\bea\label{J32}
\epsilon&=&\epsilon_1\;\epsilon_2\;,\nonumber\\
\epsilon^{\;\prime}&=&\epsilon^{\;\prime}_1=\epsilon^{\;\prime\prime}_1\;\epsilon^{\;\prime}_2\;,\nonumber\\
\epsilon^{\;\prime\prime}&=&\epsilon^{\;\prime\prime}_1\;\epsilon^{\;\prime\prime}_2\;,
\eea
and the second condition is not satisfied. If we keep the same Dirac operator (\ref{J29}), it is the definition of $\J$ that  should be changed\footnote{A general examination of the $\epsilon$ sign table and the product of two real spectral triples is done in \cite{Van}.} to 
\be\label{J33}
\J=\J_1\otimes(\J_2\;\chi_2)\;,\ee
and with this $\J$ the consistency conditions become 
\bea\label{J34}
\epsilon&=&\epsilon_1\;\epsilon_2\;\epsilon_2^{\;\prime\prime},\nonumber\\
\epsilon^{\;\prime}&=&\epsilon^{\;\prime}_1=-\epsilon^{\;\prime\prime}_1\;\epsilon^{\;\prime}_2\;,\nonumber\\
\epsilon^{\;\prime\prime}&=&\epsilon^{\;\prime\prime}_1\;\epsilon^{\;\prime\prime}_2\;,
\eea
and these are satisfied. In the rest of this section, we shall assume that these choices are made. Also, in order to simplify the forthcoming formulae, we take $\alpha=-1$ so that 
$\Fu_1=\left(\begin{array}{cc} 0 & 1\\ -1 & 0 \end{array}\right)$ and $a_{(+s)}=a_{(-s)}=+1$ which imply that 
\(\J_{1(-)}=-{\J_{1(+)}}^+=\J_{1(+)}\) and $\Di_{1(\pm)}=\Di_{(\pm s)}$. The change of $\J_2$ to $\J_2\chi_2$ will not change the representation $\pi_2^{o}$ since $\pi_2$ is even with respect to $\chi_2$.\\
The $S^0$-real structure implies that $\pi$ and $\pi^{o}$, $\Di$ and $\chi$ are block diagonal in the decomposition 
(\ref{J27}) of $\Hi$ and we obtain 
\begin{description} 
\item[the representations :] let $f\in\Fu(S^2,\Co\oplus\Co)$, then 
\be\label{J35}
\pi(f)=\pi_{(+)}(f)\oplus\pi_{(-)}(f)\;;\;\pi^{o}(f)=\pi^{o}_{(+)}(f)\oplus\pi^{o}_{(-)}(f)\;,\ee
\bea
\pi_{(\pm)}(f(x))& = &
\left(\begin{array}{cccc}
f_a(x)\Id_N & 0 & 0 & 0 \\
0 & f_a(x)\Id_N & 0 & 0 \\
0 & 0 & f_b(x)\Id_N & 0 \\
0 & 0 & 0 & f_b(x)\Id_N
\end{array}\right)\;,\nonumber\\
&&\nonumber\\
\pi^{o}_{(\pm)}(f(x))& = &
\left(\begin{array}{cccc}
f_a(x)\Id_N & 0 & 0 & 0 \\
0 & f_b(x)\Id_N & 0 & 0 \\
0 & 0 & f_a(x)\Id_N & 0 \\
0 & 0 & 0 & f_b(x)\Id_N
\end{array}\right)\;.\nonumber\\
&&\label{J36}\eea
\item[the Dirac operator :] \(\Di=\Di_{(+)}\oplus\Di_{(-)}\) is given by 
\begin{itemize}
\item in case {\bf 2.a)} :
\be\label{J37}
\Di_{{\bf a}(\pm)}=\left(\begin{array}{cccc}
\Di_{1(\pm)}\Id_N & 0 & 0 & 0 \\
0 & \Di_{1(\pm)}\Id_N& 0 & \gamma_{3}A^+ \\
0 & 0 & \Di_{1(\pm)}\Id_N & \gamma_{3}B^+ \\
0 & \gamma_{3}A & \gamma_{3} B & \Di_{1(\pm)}\Id_N \end{array}\right)\;,\ee
%%%%%%%
\item in case {\bf 2.b)} :
\be\label{J38}
\Di_{{\bf b}(\pm)}=\left(\begin{array}{cccc}
\Di_{1(\pm)}\Id_N & \gamma_{3}B^+ & \gamma_{3}A^+ & 0 \\
\gamma_{3}B & \Di_{1(\pm)}\Id_N& 0 & 0 \\
\gamma_{3}A & 0 & \Di_{1(\pm)}\Id_N & 0 \\
0 & 0 & 0 & \Di_{1(\pm)}\Id_N \end{array}\right)\;.\ee
%%%%%
\end{itemize}
\item[the chirality :] \(\chi=\chi_{(+)}\oplus\chi_{(-)}\), is given by 
\begin{itemize}
\item in case {\bf 2.a)} :
\be\label{J39}
\chi_{(\pm)}=\chi^{\;\prime}\left(\begin{array}{cccc}
\gamma_3\Id_N & 0 & 0 & 0 \\
0 & \gamma_3\Id_N & 0 & 0 \\
0 & 0 & \gamma_3\Id_N & 0 \\
0 & 0 & 0 & -\gamma_3\Id_N \end{array}\right)\;,\ee
%%%%%%%
\item in case {\bf 2.b)} :
\be\label{J40}
\chi_{(\pm)}=\chi^{\;\prime}\left(\begin{array}{cccc}
-\gamma_3\Id_N & 0 & 0 & 0 \\
0 & \gamma_3\Id_N & 0 & 0 \\
0 & 0 & \gamma_3\Id_N & 0 \\
0 & 0 & 0 & \gamma_3\Id_N \end{array}\right)\;,\ee
%%%%%%%%%
\end{itemize}
%%%%%%%% 
\item[the real structure :] \(\J=\left(\begin{array}{cc}
0 & \J_{(-)}\\ \J_{(+)} & 0 \end{array}\right)\), exchanges $\Hi_{(+)}$ and $\Hi_{(-)}$ and 
\(\J_{(\pm)}=\Fu_1\otimes\Fu_2\chi_2\) yields
\begin{itemize}
\item in case {\bf 2.a)} :
\be\label{J41}
\J_{{\bf a}(\pm)}=\chi^{\;\prime}\;\left(\begin{array}{cccc}
\Fu_{1}\Id_N & 0 & 0 & 0 \\
0 & 0 & \Fu_1\Id_N & 0  \\
0 & \Fu_1\Id_N & 0 & 0 \\
0 & 0 & 0 & -\Fu_1\Id_N \end{array}\right)\,\Ko\;,\ee
%%%%%%%
\item in case {\bf 2.b)} :
\be\label{J42}
\J_{{\bf b}(\pm)}=\chi^{\;\prime}\;\left(\begin{array}{cccc}
-\Fu_{1}\Id_N & 0 & 0 & 0 \\
0 & 0 & \Fu_1\Id_N & 0  \\
0 & \Fu_1\Id_N & 0 & 0 \\
0 & 0 & 0 & \Fu_1\Id_N \end{array}\right)\,\Ko\;,\ee
%%%%%
\end{itemize}
\end{description}
%%
%%%%%%%%%%%%%%%%%%%%%%%%%%%%%%%%%%
\subsection{The "Real" Yang-Mills-Higgs action}\label{RealYMH}
The representations $\pi_{(\pm)}$ of (\ref{J35}) and $\pi^{o}_{(\pm)}$ of (\ref{J36}), with the Dirac operators of 
(\ref{J37}) and (\ref{J38}), induce representations of $\Omega^\bullet(\Al)$.\\
Let $F\in\Omega^{(1)}(\Al)$, then, using the same techniques which led to (\ref{TRIP8}), we obtain in case {\bf 2.a)}:  
\[
\pi_{(\pm)}(F)=
\left(\begin{array}{cccc}
-\im\;\clif(\sigma^{(1)}_a)\Id_{N} & 0 & 0 & 0\\ 
0 & -\im\;\clif(\sigma^{(1)}_a)\Id_{N} & 0 & \sigma^{(0)}_{ab}\;\gamma_3 A^+ \\
0 & 0 & -\im\;\clif(\sigma^{(1)}_b)\Id_{N}& 0 \\
0 & \sigma^{(0)}_{ba}\;\gamma_3 A & 0 & -\im\;\clif(\sigma^{(1)}_b)\Id_{N}
\end{array}\right)\;,\]
and in case {\bf 2.b)}:
\[
\pi_{(\pm)}(F)=
\left(\begin{array}{cccc}
-\im\;\clif(\sigma^{(1)}_a)\Id_{N} & 0 & \sigma^{(0)}_{ab}\;\gamma_3 A^+ & 0\\ 
0 & -\im\;\clif(\sigma^{(1)}_a)\Id_{N} & 0 & 0 \\
\sigma^{(0)}_{ba}\;\gamma_3 A & 0 & -\im\;\clif(\sigma^{(1)}_b)\Id_{N}& 0 \\
0 & 0 & 0 & -\im\;\clif(\sigma^{(1)}_b)\Id_{N}
\end{array}\right)\;,\]
where the differential forms $\sigma^{(k)}$'s are given in (\ref{one-form}).\\ Introducing the $2N\times 2N$ matrix  
in case {\bf 2.a)} as \(M_{\bf (2.a)}=\left(\begin{array}{cc}0 & 0 \\ 0 & A\end{array}\right)\) and in case {\bf 2.b)} as  \(M_{\bf 2.b)}=\left(\begin{array}{cc}A & 0 \\ 0 & 0\end{array}\right)\), we may also write:  
\be\label{J43}
\pi_{(\pm)}(F)=
\left(\begin{array}{cc}
-\im\;\clif(\sigma^{(1)}_a)\Id_{2N} & 
\sigma^{(0)}_{ab}\;\gamma_3 M^+ \\
\sigma^{(0)}_{ba}\;\gamma_3 M & 
-\im\;\clif(\sigma^{(1)}_b)\Id_{2N}
\end{array}\right)\;.\ee 
A universal two-form $G\in\Omega^{(2)}(\Al)$ has representations $\pi_{(\pm)}(G)$ given by a similar expression as 
in (\ref{TRIP9}). The unwanted differential ideal ${\cal J}$ is removed using the orthonality condition analogous to (\ref{Nojunk})  
\be\label{J44}
\rho_{aaa}^{(0)}-{1\over 2N}\rho_{aba}^{(0)}\;{\bf tr}\{M^+M\}=0\quad;\quad
\rho_{bbb}^{(0)}-{1\over 2N}\rho_{bab}^{(0)}\;{\bf tr}\{MM^+\}=0\;.\ee
The representative of $G$ in $\Omega^{(2)}_D(\Al)$, as in (\ref{TRIP15}), is given by :
\be\label{J45}
\pi_{D(\pm)}(G)=\left(\begin{array}{cc}
\pi_{D(\pm)}(G)_{[aa]} & \pi_{D(\pm)}(G)_{[ab]}\\
\pi_{D(\pm)}(G)_{[ba]} & \pi_{D(\pm)}(G)_{[bb]}\end{array}\right)\;,\ee
with
\beann
\pi_{D(\pm)}(G)_{[aa]}&=&-\clif(\rho_{aaa}^{(2)})\otimes\Id_{2N}+ \rho_{aba}^{(0)}\otimes\Bigl[M^+M\Bigr]_{NT} \;,\\
\pi_{D(\pm)}(G)_{[ab]}&=&-\im\;\clif(\rho_{ab}^{(1)})\gamma_{3(\pm)}\otimes M^+\;,\\
\pi_{D(\pm)}(G)_{[ba]}&=&-\im\;\clif(\rho_{ba}^{(1)})\gamma_{3(\pm)}\otimes M\;,\\ 
\pi_{D(\pm)}(G)_{[bb]}&=&-\clif(\rho_{bbb}^{(2)})\otimes\Id_{2N}+\rho_{bab}^{(0)}\otimes\Bigl[MM^+\Bigr]_{NT}\;,
\eeann
where the differential forms $\rho^{(k)}$ are given in (\ref{two-form}) and 
\beann
\Bigl[M^+M\Bigr]_{NT}&=& M^+M-{1\over 2N}{\bf tr}\{M^+M\}\;,\\
\Bigl[M^+M\Bigr]_{NT}&=& MM^+-{1\over 2N}{\bf tr}\{MM^+\}\;.\eeann
The scalar product in $\Omega^{(2)}_D(\Al)$ is the same for representatives in $\Hi_{(+)}$ or $\Hi_{(-)}$ and is given by a similar expression as (\ref{TRIP16}) :
\bea\label{J46}
&&\langle \pi_{D(\pm)}(G);\pi_{D(\pm)}(G^\prime)\rangle_{2,D}=\nonumber\\
&&{1\over 2\pi}
\Biggl\{
2N\;\int_{S^2}\left({\rho^{(2)}_{aaa}}^*\wedge\star\rho^{\prime(2)}_{aaa}
+{\rho^{(2)}_{bbb}}^*\wedge\star\rho^{\prime(2)}_{bbb}\right)\nonumber\\
&& 
+{\bf tr}\{A^+A\}\;
\int_{S^2}\left({\rho^{(1)}_{ab}}^*\wedge\star\rho^{\prime(1)}_{ab}
+
{\rho^{(1)}_{ba}}^*\wedge\star\rho^{\prime(1)}_{ba}\right)\nonumber\\
&&
+\left[{\bf tr}\{(A^+A)^2\}-{1\over 2N}({\bf tr}\{A^+A\})^2\right]\;
\int_{S^2}\left({\rho^{(0)}_{aba}}^*\wedge\star\rho^{\prime(0)}_{aba}
+
{\rho^{(0)}_{bab}}^*\wedge\star\rho^{\prime(0)}_{bab}\right)\Biggr\}\;.\nonumber\\
&&
\eea
From this expression of the scalar product, the Yang-Mills-Higgs action is essentially twice the action (\ref{YMH7}) obtained in section {\bf \ref{Yang-Mills-Higgs}} : 
%%%%%%
\bea\label{J47}
&&{\bf S}_{YMH}(\nabla_D)={\lambda\over \pi}\biggl\{\nonumber\\
&& 2N\;\int_{S^2}\left({\bf tr}_{matrix}\Bigl\{\fatl F_a\fatr^+\wedge\star\fatl F_a\fatr\Bigr\}
+F_b^*\wedge\star F_b\right)\nonumber\\
&& +2\;{\bf tr}\{A^+A\}\;
\int_{S^2}\langle\nabla\eta_{ba}\vert\wedge\star\vert\nabla\eta_{ab}\rangle\nonumber\\
&& +\left[{\bf tr}\{(A^+A)^2\}-{1\over 2N}({\bf tr}\{A^+A\})^2\right]\;
\int_{S^2}\star\Bigl(2(\langle \eta_{ba}\vert \eta_{ab}\rangle-1)^2+1\Bigr)\biggr\}\;.\nonumber\\
&&\eea
%%%%%%%%%%%%%%%%%%
%%%%%%%%%%%%%%%%%%
\subsection{The "Real" covariant Dirac operator}
\label{RealMatter}
%%%%%%%%%
Matter\footnote{In this section we consider case {\bf 2.a)} only. At the end the final result 
for the covariant Dirac operator in case {\bf 2.b)} will also be given} in this "real spectral
triple" approach is represented by states of the covariant Hilbert space
$\Hi_{Cov}=\Mod\otimes_\Al\Hi\otimes_\Al\Mod^*$. Fot the $S^0$-real spectral triple
$\Hi=\Hi_{(+)}\oplus\Hi_{(-)}$ so that also $\Hi_{Cov}$ splits in a sum of "particle" and
"antiparticle" Hilbert spaces $\Hi_{Cov}=\Hi_{(+{\bf p})}\oplus\Hi_{(-{\bf p})}$ where each 
$\Hi_{(\pm{\bf p})}=\Mod\otimes_\Al\Hi_{(\pm)}\otimes_\Al\Mod^*$ has  typical elements  
\(\Vert\BPsi_{(\pm{\bf p}}\BRR\). The projective module $\Mod =\Proj \Al^2$ and its dual $\Mod^*$
were examined in section {\bf \ref{Module}} and in appendix {\bf \ref{ModCon}}. In bases $\{E_i\}$
and $\{E^j\}$ of the free modules $\Al^2$ and ${\Al^2}^*$, we may represent a state of
$\Al^2\otimes_\Al\Hi_{(\pm)}\otimes_\Al{\Al^2}^*$ as 
\(E_i\otimes_\Al{\Psi_{(\pm)}}^i_{\;j}\otimes_\Al E^j\). It is a state 
\(\Vert\BPsi_{(\pm{\bf p}}\BRR\) of $\Hi_{(\pm{\bf p})}$ if ${\Psi_{(\pm)}}^i_{\;j}\in \Hi_{(\pm)}$
obeys
\be\label{J48}
{\Psi_{(\pm)}}^i_{\;j}=\pi_{(\pm)}(P^i_k)\pi_{(\pm)}^o(P^\ell_j)\,{\Psi_{(\pm)}}^k_{\;\ell}\;.\ee
Since $P^i_{\;k,a}(x)=\delta^i_k$ and $P^i_{\;k,b}(x)=\vert\nu(x)\rangle^i\langle\nu(x)\vert_k$ 
(cf. section {\bf \ref{Module}}) the vector ${\Psi_{(\pm)}}^i_{\;j}\in\Hi_{(\pm)}$ is represented
by the column vector
\be\label{J49}
{\Psi_{(\pm)}}^i_{\;j}=\left(\begin{array}{c}
{\fatl\psi_{(\pm)aa}\fatr}^i_{\;j}\\ 
{\vert\psi_{(\pm)ab}\rangle}^i\langle\nu\vert_j\\ \vert\nu\rangle^i{\langle\psi_{(\pm)ba}\vert}_j\\
\vert\nu\rangle^i\;{\psi_{(\pm)bb}}\;\langle\nu\vert_j\end{array}\right)\;,\ee
where $\fatl\psi_{(\pm)aa}\fatr$ is a quadruplet and 
$\psi_{(\pm)bb}=\langle\nu\vert\fatl\psi_{(\pm)bb}\fatr\vert\nu\rangle$ a singlet of Pensov spinor
fields of spin weight $\pm s$, while
$\vert\psi_{(\pm)ab}\rangle=\fatl\psi_{(\pm)ab}\fatr\vert\nu\rangle$, respectively
$\langle\psi_{(\pm)ba}\vert=\langle\nu\vert\fatl\psi_{(\pm)ba}\fatr$, are doublets of Pensov
spinors of weight $(\pm s)-n/2$, respectively $(\pm s)+n/2$.\\ The covariant real structure
$\J_{Cov}$, as defined in appendix {\bf \ref{RealCov}}, acts on $\Vert\BPsi_{Cov}\BRR$ as 
\be\label{J50}
{\left(\J_{Cov}\Vert\BPsi_{Cov}\BRR\right)_{(\pm)}}^i_{\;j}=\chi^{\;\prime}
\left(\begin{array}{c}
\delta^{i\bar\ell}\left(\Fu_1\Ko\fatl\psi_{(\mp)aa}\fatr^k_{\;\ell}\right)\delta_{\bar kj}\\
\delta^{i\bar\ell}\left(\Fu_1\Ko\langle\psi_{(\mp)ba}\vert_\ell\right)\langle\nu\vert_j\\
\vert\nu\rangle^i\left(\Fu_1\Ko\vert\psi_{(\mp)ab}\rangle^k\right)\delta_{\bar kj}\\
-\vert\nu\rangle^i\left(\Fu_1\Ko\psi_{(\mp)bb}\right)\langle\nu\vert_j\end{array}\right)\;.\ee
\newpage
\noindent
The covariant Dirac operator, defined in (\ref{Cov10}), is also block diagonal : $\Di_\nabla=\Di_{(+{\bf p})}\oplus\Di_{(-{\bf p})}$ and is given by :
\bea\label{J51}
&&\Di_{(\pm{\bf p})}\biggl\{(E_i P^i_{\;k})\otimes_\Al{\psi_{(\pm)}}^k_{\;\ell}\otimes_\Al (P^\ell_{\;j}E^j)\biggr\}
=\nonumber\\
&&
E_i\otimes_\Al\pi_{(\pm)}\left(\fatl A\fatr^i_{\;k}\right){\psi_{(\pm)}}^k_{\;j}
\otimes_\Al E^j\nonumber\\
&&+E_i\otimes_\Al
\pi_{(\pm)}(P^i_{\;k})\pi_{(\pm)}^o(P^\ell_j)\Di_{(\pm)}{\psi_{(\pm)}}^k_{\;\ell}
\otimes_\Al E^j\nonumber\\
&&+\epsilon^{\;\prime}E_i\otimes_\Al 
\pi_{(\pm)}^o\left(\fatl A\fatr^\ell_j\right){\psi_{(\pm)}}^i_{\;\ell}
\otimes_\Al E^j\;,\eea
where $\fatl A\fatr$ is the $2\times 2$ matrix of universal one-forms given in (\ref{PMod6}).\\
It is represented by the matrix valued differential one- and zero-forms given in terms of \(\fatl\alpha_a\fatr\), \(\alpha_b\), \(\vert\Phi_{ab}\rangle\) and \(\langle\Phi_{ba}\vert\), defined in 
(\ref{YMH1}) by :
\bea\label{J52}
\fatl A_{a}\fatr&=&-\im\gamma^r\fatl\alpha_{a,r}\fatr\;,\nonumber\\
\fatl A_{b}\fatr&=&-\im\gamma^r\alpha_{b,r}\vert\nu\rangle\langle\nu\vert\;,\nonumber\\
\fatl A_{ab}\fatr&=&\vert\Phi_{ab}\rangle\langle\nu\vert\;\gamma_{3}\;,\nonumber\\
\fatl A_{ba}\fatr&=&\vert\nu\rangle\langle\Phi_{ba}\vert\;\gamma_{3}\;.
\eea
We obtain :
\be\label{J53}
\pi_{(\pm)}(\fatl A\fatr)=
\left(\begin{array}{cccc}
\fatl A_{a}\fatr\Id_N & 0 & 0 & 0 \\
0 & \fatl A_{a}\fatr\Id_N & 0 & \fatl A_{ab}\fatr {\bf A}^+ \\
0 & 0 & \fatl A_{b}\fatr\Id_N & 0 \\
0 & \fatl A_{ba}\fatr {\bf A} & 0 & \fatl A_{b}\fatr\Id_N
\end{array}\right)\,.\ee 
The $\pi_{(\pm)}^o$ representative of $\fatl A\fatr$ is computed as :
\be\label{J54}
\epsilon^{\;\prime}\pi_{(\pm)}^o(\fatl A\fatr)=
\left(\begin{array}{cccc}
\fatl A_{a}\fatr\Id_N & 0 & 0 & 0 \\
0 & \fatl A_{b}\fatr\Id_N & 0 & 0 \\
0 & 0 & \fatl A_{a}\fatr\Id_N & -\fatl A_{ba}\fatr {\bf B}^+  \\
0 & 0 & -\fatl A_{ab}\fatr {\bf B} & \fatl A_{b}\fatr\Id_N
\end{array}\right)\;.\ee 
\newpage
\noindent
Substituting (\ref{J53}) and (\ref{J54}) in (\ref{J51}), we obtain :
\bea
\fatl\Di_\nabla\Vert\BPsi_{Cov}\BRR_{(\pm)aa}\fatr&=&
\Di_{1(\pm)}\fatl\psi_{(\pm)aa}\fatr -\im\;\gamma^r\left[\fatl\alpha_{a,r}\fatr,\fatl\psi_{(\pm)aa}\fatr\right]\;,\nonumber\\
\vert\Di_\nabla\Vert\BPsi_{Cov}\BRR_{(\pm)ab}\rangle&=&
\Di^{(-n/2)}_{1(\pm)} \vert\psi_{(\pm)ab}\rangle+\vert\eta_{ab}\rangle\gamma_{3}{\bf A}^+\psi_{(\pm)bb}\nonumber\\
&&-\im\;\gamma^r\left(\fatl\alpha_{a,r}\fatr\vert\psi_{(\pm)ab}\rangle
-\vert\psi_{(\pm)ab}\rangle\alpha_{b,r}\right)\;;\nonumber\\
\langle\Di_\nabla\Vert\BPsi_{Cov}\BRR_{(\pm)ba}\vert&=& 
\Di^{(+n/2)}_{1(\pm)}\langle\psi_{(\pm)ba}\vert+\langle\eta_{ba}\vert\gamma_{3}{\bf B}^+\psi_{(\pm)bb}\nonumber\\
&&-\im\;\gamma^r\left(\alpha_{b,r}\langle\psi_{(\pm)ba}\vert 
-\langle\psi_{(\pm)ba}\vert\fatl\alpha_{a,r}\fatr\right)\;,\nonumber\\
\Di_\nabla\Vert\BPsi_{Cov}\BRR_{(\pm)bb}&=&
\Di_{1(\pm)}\psi_{(\pm)bb}\nonumber\\
&&+\gamma_{3}{\bf A}\langle\eta_{ba}\vert\psi_{(\pm)ab}\rangle +
\gamma_{3}{\bf B}\langle\psi_{(\pm)ba}\vert\eta_{ab}\rangle\;.\nonumber\\
&&\label{J55}
\eea
In case {\bf 2.b)} a similar result is obtained :
\bea
\fatl\Di_\nabla\Vert\BPsi_{Cov}\BRR_{(\pm)aa}\fatr&=&
\Di_{1(\pm)}\fatl\psi_{(\pm)aa}\fatr -\im\;\gamma^r\left[\fatl\alpha_{a,r}\fatr,\fatl\psi_{(\pm)aa}\fatr\right]\nonumber\\
&&+\gamma_{3}{\bf A}^+\;\vert\eta_{ab}\rangle\langle\psi_{(\pm)ba}\vert
+\gamma_{3}{\bf B}^+ \;\vert\psi_{(\pm)ab}\rangle\langle\eta_{ba}\vert\;, \nonumber\\
\vert\Di_\nabla\Vert\BPsi_{Cov}\BRR_{(\pm)ab}\rangle&=& 
\Di^{(-n/2)}_{1(\pm)}\vert\psi_{(\pm)ab}\rangle
+\gamma_{3}{\bf B}\; \fatl\psi_{(\pm)aa}\fatr\vert\eta_{ab}\rangle\;,\nonumber\\
&&-\im\;\gamma^r\left(\fatl\alpha_{a,r}\fatr\vert\psi_{(\pm)ab}\rangle
-\vert\psi_{(\pm)ab}\rangle\alpha_{b,r}\right)\;,\nonumber\\ 
\langle\Di_\nabla\Vert\BPsi_{Cov}\BRR_{(\pm)ba}\vert&=& 
\Di^{(+n/2)}_{1(\pm)}\langle\psi_{(\pm)ba}\vert
+\gamma_{3}{\bf A} \;\langle\eta_{ba}\vert\fatl\psi_{(\pm)aa}\fatr\;,\nonumber\\
&&-\im\;\gamma^r\left(\alpha_{b}\langle\psi_{(\pm)ba}\vert-\langle\psi_{(\pm)ba}\vert\fatl\alpha_{a,r}\fatr\right)\nonumber\\
\left(\Di_\nabla\Vert\BPsi_{Cov}\BRR\right)_{(\pm)bb}&=&
\Di_{1(\pm)}\psi_{(\pm)bb}\;.\nonumber\\
&&\label{J56}
\eea
The difference is that here the Higgs field interact with the quadruplet $\fatl\psi_{(\pm)aa}\fatr$, while in case {\bf 2.a)} it interacts with the singlet $\psi_{(\pm)bb}$ and it is this interaction that gives masses to the particles.\\ The Dirac operators,  
$\Di^{(-n/2)}_{1(\pm)}=\Di_{(\pm s-n/2)}$ and $\Di^{(+n/2)}_{1(\pm)}=\Di_{(\pm s+n/2)}$,
acting on Pensov spinor fields of spin weight $\pm s+n/2$ or $\pm s-n/2$, arise from   
\[\Bigl(\Di_{1(\pm)}(\vert\psi_{(\pm)ab}\rangle\langle\nu\vert)\Bigr)\vert\nu\rangle=
\Di^{(-n/2)}_{1(\pm)}\vert\psi_{(\pm)ab}\rangle\;,\]
\[\langle\nu\vert\Bigl(\Di_{1(\pm)}(\vert\nu\rangle\langle\psi_{(\pm)ba}\vert)\Bigr)=
\Di^{(+n/2)}_{1(\pm)}\langle\psi_{(\pm)ba}\vert\;,\]
where $\fatl P_b\fatr=\vert\nu\rangle\langle\nu\vert$ is the representative, chosen in (\ref{PMod3}), of the homotopy class $[n]\in\pi_2(S^2)$.
The induced contribution of the "magnetic monopole" is hidden in this modification of the Dirac operator.\\ The Higgs doublets of Pensov fields of weight $\mp \;n/2$, $\vert\eta_{ab}\rangle$ and $\langle\eta_{ba}\vert$ were  defined in (\ref{YMH3}).\\
A suitable action of the matter field would be 
\be\label{J59}
{\bf S}_{\bf Mat}(\Vert\BPsi_{Cov}\BRR,\nabla_D)=
\Bigl(\Vert\BPsi_{Cov}\BRR\;;\;\Dirac_\nabla\Vert\BPsi_{Cov}\BRR\Bigr)\;.\ee
But, if we aim for a theory admitting only chiral matter, i.e. if we restrict the Hilbert space to those vectors obeying say 
\be\label{J60}
\chiral\Vert\BPsi_{Cov}\BRR=+\Vert\BPsi_{Cov}\BRR\;,\ee
then the above action vanishes identically. Another choice for the action would be 
\be\label{J61}
{\bf S}_{\bf Chiral}(\Vert\BPsi_{Cov}\BRR,\nabla_D)=
\Bigl(\J_{Cov}\Vert\BPsi_{Cov}\BRR\;;\;\Dirac_\nabla\Vert\BPsi_{Cov}\BRR\Bigr)\;.\ee
It is easy to show that this action does not vanish identically if 
\be\label{J62}
\epsilon^{\;\prime\prime}=-1\;,\ee
\be\label{J63}
\epsilon\;\epsilon^{\;\prime}=+1\;.\ee
In two dimensions Connes' sign table obeys the first but not the second 
condition\footnote{In four dimensions it is the first condition that is not satisfied.}. It should
however be stressed that Connes' sign table, with its modulo eight periodicity, comes from
representation theory of the real Clifford algebras and if we restrict our (generalized) spinors to
Weyl spinors, we loose the Clifford {\it algebra} representation and the sign tables ceases to be
mandatory.  We could then go back to {\bf \ref{Prodtriple}} and 
\begin{itemize}
\item with $\J=\J_1\otimes\J_2$ require (\ref{J32}) to hold with 
$\epsilon_1^{\;\prime}=-1\epsilon_1$ or,
\item with $\J=\J_1\otimes(\J_2\chi_2)$ require (\ref{J34}) to hold with 
$\epsilon_1^{\;\prime}=+1=\epsilon_1$.
\end{itemize}
%The full story of this chirality problem, sometimes called the "neutrino paradigm", is postponed 
%to forthcoming work on the four-sphere with $\Co\oplus{\bf H}$.
%%%%%%%%%%%%
\newpage
%%%%%%%%%%%%
\section{Conclusions and Outlook}
\label{Out}
The Connes-Lott model over the two-sphere, with $\Co\oplus\Co$ as discrete algebra, has been
generalized such as to allow for a nontrivial topological structure. The basic Hilbert space
$\Hi_{(s)}$ was made of Pensov spinors which can be interpreted as usual spinors inteacting with a
Dirac monopole "inside" the sphere. Covariantisation of the Hilbert space
$\Hi_{(s)}\otimes\Hi_{dis}$ with a nontrivial projective module $\Mod$ induced a "spin" change in
certain matter fields so that we obtained singlets and doublets of different spin content. The Higgs
fields also acquired a nontrivial topology since they are no longer ordinary functions on
the sphere, but rather Pensov scalars i.e. sections of nontrivial line bundles over the sphere.\\
A real spectral triple has also been constructed essentially through the doubling of the Pensov
spinors so that the Hilbert space of the continuum spectral triple
became $\Hi_1=\Hi_{(s)}\oplus\Hi_{(-s)}$. The discrete spectral triple had also to be extended in
order that the first order condition could be met.
In contrast with the standard noncommutative geometry model of the standard model, in our model the continuum spectral triple has an $S^0$-real structure while the discrete spectral triple has not. Some physical plausability arguments for this were given in section {\bf \ref{RPDtriple}}. It was also shown that the covariantisation of the real spectral triple with the nontrivial $\Mod$ allows the abelian gauge fields to survive, while they are slain if covaraiantisation is done with a trivial module. Finally a possibility of solution to the the problem of a non vanishing action of chiral matter has been indicated, paying the price of using a complex action.\\
If we address the quantisation problem in a path integral formalism, let us first recall that we have Higgs fields which are Pensov scalars of a $\Pe^{(s_1)}$ and matter fields Pensov spinors of type $\Hi_{(s_2)}$. It is thus tempting to assume that the Higgs fields should be even Grassmann variables is $s_1$ is integer valued and odd Grassmann variables if it is half-integer. In the same vein the matter fields should be odd or even Grassmann variables if $s_2$ is integer or half integer valued. A thorough examination of this issue is however beyond the scope of this work.
\newpage
%%%%%%%%%
\appendix 
\section{On modules and connections}
\label{ModCon}\setcounter{equation}{0} 
%%%%%%%
Let $\Mod$ be a right module over the $\star$-algebra $\Al$. The set of endomorphisms $\End=END_\Al(\Mod)$ has an obvious algebra structure and he left action of ${\bf A}\in\End$ on $X\in\Mod$ commutes with the right action of $a\in\Al$ :
\(({\bf A}X)a={\bf A}(Xa)\) so that $\Mod$ acquires canonicall a $\End -\Al$ bimodule structure. The dual module $\Mod^*=HOM_\Al(\Mod,\Al)$ is a left $\Al$-module and ${\bf A}\in \End$ acts on the right on $\xi\in\Mod^*$ by \((\xi{\bf A},X)=(\xi,{\bf A}X)\) so that $\Mod^*$ is a $\Al-\End$ bimodule.\\ When $\Mod$ is endowed with a sesquilinear, hermitian and non-degenerate form \({\bf h}:\Mod\times\Mod\rightarrow\Al:(X,Y)\rightarrow{\bf h}(X,Y)\), there is a canonical bijective mapping 
\be\label{MC1}
{\bf h}^\sharp:\Mod\rightarrow\Mod^*: X\rightarrow{\bf h}^\sharp(X)\vdash\left({\bf h}^\sharp(X),Y\right)={\bf h}(X,Y)\;.\ee 
Since ${\bf h}^\sharp(Xa)=a^*{\bf h}^\sharp(X)$, ${\bf h}^\sharp$ is an anti-isomorphism with inverse 
\[({\bf h}^\sharp)^{-1}={\bf h}^\flat :\xi\rightarrow{\bf h}^\flat(\xi)\;,\]
and \({\bf h}^\flat(a\xi)={\bf h}^\flat(\xi)a^*\).
The inverse form ${\bf h}^{-1}$ is defined by :
\[{\bf h}^{-1}:\Mod^*\times\Mod^*\rightarrow\Al:(\xi,\eta)\rightarrow{\bf h}^{-1}(\xi,\eta)={\bf h}\left({\bf h}^\flat(\xi),{\bf h}^\flat(\eta)\right)\;.\]
The hermitian conjugate of ${\bf A}\in\End$ is defined by \({\bf h}(X,{\bf A}^+Y)={\bf h}({\bf A}X,Y)\).\\
Let $\Omega^\bullet(\Al)=\bigoplus_{k\in{\bf Z}}\Omega^{(k)}(\Al)$ be a graded differential $\ast$-envelope of $\Al$, then ${\bf h}$ can be extended to $\Mod^\bullet=\Mod\otimes_\Al\Omega^\bullet(\Al)$ by :
\be\label{MC2}
{\bf h}(X\otimes_\Al F,Y\otimes_\Al G)= F^+{\bf h}(X,Y) G\;,\ee
where $X,Y\in\Mod$ and $F,G\in\Omega^\bullet(\Al)$.\\
Defining $\Mod^{*\bullet}\doteq\Omega^\bullet(\Al)\otimes_\Al\Mod^*$, ${\bf h}^\sharp$ can be extended as a mapping  
$\Mod^\bullet\rightarrow\Mod^{*\bullet}$ by $(X\otimes_\Al F)^\dagger=F^+\otimes_\Al X^\dagger$, where \({\bf h}^\sharp(\;)\) is written as $(\;)^\dagger$.\\
A connection $\nabla$ in $\Mod$ is an additive map 
\be\label{MC3}
\nabla : \Mod\rightarrow \Mod\otimes_\Al\Omega^{(1)}(\Al): X\rightarrow \nabla X\;,\ee
which is additive and obeys the Leibniz rule
\[\nabla(Xa)=(\nabla X)a+X\otimes_\Al \dif a\;.\]
It defines an associate dual connection $\nabla^*$ in $\Mod^*$ by :
\be\label{MC4}
\left(\nabla^*\xi,X\right)+\left(\xi,\nabla X\right)=\dif\left(\xi,X\right)\;.\ee
It obeys \(\nabla^*(a\xi)=\dif a\otimes_\Al\xi+a(\nabla^*\xi)\).\\
The extension of $\nabla$ to $\Mod^\bullet$ by $\nabla(X\otimes_\Al F)=(\nabla X)F+X\otimes_\Al\dif F$, allows to define the curvature as $\nabla^2:\Mod^\bullet\rightarrow\Mod^\bullet$ and it is seen that $\nabla^2$ is in fact an endomorphism of $\Mod^\bullet$ considered as a right $\Omega^\bullet$-module : 
\be\label{MC5}
\nabla^2\Bigl((X\otimes_\Al F)\;G\Bigr)=\left(\nabla^2(X\otimes_\Al F)\right)G\;.\ee
Similarly, $\nabla^*$ is extended to $\Mod^{*\bullet}$ and 
\[\left(\nabla^{*2}(G\otimes_\Al\xi),X\otimes_\Al F\right)=\left(G\otimes_\Al\xi,\nabla^2(X\otimes_\Al F)\right)\;.\]
When $\Mod$ has an hermitian structure, the connection is said to be compatible with this hermitian structure, if\footnote{Here we have chosen $\dif(a^*)=-(\dif a)^*$.}
\be\label{MC6}
\dif\left({\bf h}(X,Y)\right)=-{\bf h}(\nabla X,Y)+{\bf h}(X,\nabla Y)\;\ee
or equivalently if
\[\dif\left({\bf h}^{-1}(\xi,\eta)\right)={\bf h}^{-1}(\nabla^* \xi,\eta)-{\bf h}^{-1}(\xi,\nabla^* \eta)\;.\]
The mapping ${\bf h}^\sharp$ relates both connections through
\be\label{MC7}
\nabla^*{\bf h}^\sharp(X)=-{\bf h}^\sharp\left(\nabla X\right)\;\mbox{or}\;\nabla^*X^\dagger=-\left(\nabla X\right)^\dagger\;.\ee
The curvature of a compatible connection is hermitian :
\be\label{MC8}
{\bf h}(\nabla^2 X,Y)={\bf h}(X,\nabla^2 Y)\;.\ee
The algebras $\Al$ and $\End=END_\Al(\Mod)$ are said to be Morita equivalent in the sense that there exists a $\End-\Al$ bimodule $\Mod$ and a $\Al-\End$ bimodule $\Mod^*$ such that $\Mod\otimes_\Al\Mod^*\simeq\End$, with the identification 
\((X\otimes_\Al\xi)\;Y=X\;(\xi,Y)\) and \(\eta\;(X\otimes_\Al\xi)=(\eta,X)\;\xi\), and $\Mod^*\otimes_\End\Mod\simeq\Al$ with \(Y\;(\xi\otimes_\End X)=Y\;(\xi,X)\) and \((\xi\otimes_\End X)\;\eta=(\xi,X)\;\eta\).
%%%
\newpage
%%%%%%%%%
\section{Real Covariant Spectral Triples}
\label{RealCov}\setcounter{equation}{0} 
%%%%%%%
Let $\left\{\Al,\Hi,\Di,\chi,\J\right\}$ be a real spectral triple with a faithful *-representation in the bounded operators of $\Hi$, $\pi:\Al\rightarrow\End(\Hi):a\rightarrow\pi(a)$. The Dirac operator $\Di$ allows to extend this *-representation  to $\Omega^\bullet(\Al)$ by 
\be\label{Cov1}
\pi:\Al\rightarrow\End(\Hi):a_0\dif a_1\cdots\dif a_k\rightarrow\pi(a_0)[\Di,\pi(a_1)]\cdots[\Di,\pi(a_k)]\;.\ee
The real structure $\J$ defines a right action of $\Al$ on $\Hi$ or equivalently a representation of the opposite algebra $\Al^o$ by
\[\pi^o:\Al^o\rightarrow\End(\Hi):a\rightarrow\pi^o(a)=\J\left(\pi(a)\right)^+\J^+\;.\]
It is also generalized to $\Omega^\bullet(\Al)$ by\footnote{This definition reads 
\(\pi^o\Bigl(a_0\dif a_1\cdots\dif a_k\Bigr)=[\pi^o(a_k),\Di^{\;\prime}]\cdots[\pi^o(a_1),\Di^{\;\prime}]\pi^o(a_0)\), where $\Di^{\;\prime}=\J\Di\J^+=\epsilon^{\;\prime}\Di$.}:
\be\label{Cov2}
\pi^o:\Al^o\rightarrow\End(\Hi):F\rightarrow\pi^o(F)=\J\left(\pi(F)\right)^+\J^+\;.\ee
It is asumed that $[\pi(a),\pi^o(b)]=0$ so that the Hilbert space $\Hi$ is provided with a $\Al$-bimodule structure and the tensor product 
\be\label{Cov3}
\Hi_{Cov}=\Mod\otimes_\Al\Hi\otimes_\Al\Mod^*\;\ee
is well defined. The vectors $\Vert\;\BRR$ of $\Hi_{Cov}$ are generated by elements of the form\footnote{ Elements of $\Hi_{Cov}$ could also be represented by $X\otimes_\Al|\Psi\rangle\otimes_\Al Y^\dagger$, if we choose to represent $\eta\in\Mod^*$ as $Y^\dagger$.}
$X\otimes_\Al|\Psi\rangle\otimes_\Al \eta$.
A scalar product $\Bigl(\Vert\;\BRR\;;\;\Vert\;\BRR\Bigr)$ in $\Hi_{Cov}$ is defined in terms of the scalar product 
$\Bigl(|\;\rangle;|\;\rangle\Bigr)$ in $\Hi$, by 
\be\label{Cov4}
\Bigl(X\otimes_\Al|\psi\rangle\otimes_\Al \eta\;\|\;R\otimes_\Al|\phi\rangle\otimes_\Al \sigma\Bigr) =
\Bigl(|\psi\rangle;\pi({\bf h}(X,R))\pi^o({\bf h}^{-1}(\sigma,\eta))|\phi\rangle\Bigr)\;,\ee
where $X,R\in\Mod$, $\eta,\sigma\in\Mod^*$, and $|\psi\rangle,|\phi\rangle\in\Hi$.\\
It is well defined and, as usual, completion of $\Hi_{Cov}$ defines the covariant Hilbert space also denoted by $\Hi_{Cov}$.\\
The representation $\pi$ defines a mapping \(\tilde\pi\) from $\Mod^\bullet$ to the bounded operators \(\End(\Hi,\Mod\otimes_\Al\Hi)\) by :
\be\label{Cov5}
\tilde\pi(X\otimes_\Al F)|\psi\rangle=X\otimes_\Al\pi(F)|\psi\rangle\;.\ee
It obeys $\tilde\pi(X\otimes_\Al FG)=\tilde\pi(X\otimes_\Al F)\circ\pi(G)$.\\
In the same way, $\pi^o$ defines a mapping \(\tilde\pi^o\) from $\Mod^{*\bullet}$ to the bounded operators \(\End(\Hi,\Hi\otimes_\Al\Mod^*)\) by :
\be\label{Cov6}
\tilde\pi^o(G\otimes_\Al \eta)|\phi\rangle=\pi^o(G)|\phi\rangle\otimes_\Al \eta\;,\ee
and $\tilde\pi^o(FG\otimes_\Al \eta)=\tilde\pi^o(G\otimes_\Al \eta)\circ\pi^o(F)$.\\
It is further assumed that $\Di$ is a first-order operator, i.e. $\forall a,b\in\Al$,
\be\label{Cov7}
\Bigl[[\Di,\pi(a)],\pi^o(b)\Bigr]=0\;.\ee
Since $[\pi(a),\pi^o(b)]=0$, it follows also that $\Bigl[[\Di,\pi^o(b)],\pi(a)\Bigr]=0$ and more generally that :
\be\label{Cov8}
[\pi(F),\pi^o(b)]=0\;;\;[\pi^o(G),\pi(a)]=0\;.\ee
It follows that
\bea
\tilde\pi(x\otimes_\Al F)|\psi\rangle\otimes_\Al b\eta &=&
\tilde\pi(X\otimes_\Al F)\Bigl(\pi^o(b)|\psi\rangle\Bigr)\otimes_\Al \eta\nonumber\\
(Xa)\otimes_\Al\tilde\pi^o(G\otimes_\Al \eta)|\psi\rangle&=&
X\otimes_\Al\tilde\pi^o(G\otimes \eta)\Bigl(\pi(a)|\psi\rangle\Bigr)\;.\label{Cov9}\eea
The covariant Dirac operator in $\Hi_{Cov}$ is defined by\footnote{The $\epsilon^{\;\prime}$ sign appears due to our definition (\ref{Cov2}).} :
\bea
\Di_{\nabla}\Bigl(X\otimes_\Al|\psi\rangle\otimes_\Al \eta\Bigl)&=&
\tilde\pi(\nabla X)|\psi\rangle\otimes_\Al \eta\nonumber\\
&& +X\otimes_\Al\Di|\psi\rangle\otimes_\Al \eta\nonumber\\
&& -\epsilon^{\;\prime}X\otimes_\Al\tilde\pi^o(\nabla^*\eta)|\psi\rangle\;.\label{Cov10}\eea
As can be checked using (\ref{Cov8}) and (\ref{Cov9}), this operator is well defined on the tensor product over $\Al$ defining $\Hi_{Cov}$.\\
The chirality in $\Hi_{Cov}$ is defined as 
\be\label{Cov11}
\chiral\Bigl(X\otimes_\Al|\psi\rangle\otimes_\Al \eta\Bigr)= X\otimes_\Al\chi|\psi\rangle\otimes_\Al \eta\;.\ee
Since $\chi$ commutes with $\pi(a)$ and $\pi^o(a)$ and anticommutes with $\Di$, it follows that 
\[\chiral\pi(F)=\pi(\underline{\alpha}(F))\chiral\;;\;\chiral\pi^o(F)=\pi^o(\underline{\alpha}(F))\chiral\;,\]
where $\underline{\alpha}$ is the main automorphism in the graded algebra $\Omega^\bullet(\Al)$.
It is then easy to show that 
\be\label{Cov12}
\chiral\Di_{\nabla}+\Di_{\nabla}\chiral=0\;.\ee
The real structure in $\Hi_{Cov}$ is defined by :
\be\label{Cov13}
\J_{\bf h}\Bigl(X\otimes_\Al|\psi\rangle\otimes_\Al \eta\Bigr)={\bf h}^\flat(\eta)\otimes_\Al\J|\psi\rangle\otimes_\Al {\bf h}^\sharp(X)\;.\ee
Obviously, if $\J^2=\epsilon\Id$, then also $\J_{\bf h}^2=\epsilon\Id$ and,\\
if $\J\chi=\epsilon^{\;\prime\prime}\chi\J$, also
$\J_{\bf h}\chiral=\epsilon^{\;\prime\prime}\chiral\J_{\bf h}$ holds.\\
Using
\beann
\J_{\bf h}\Bigl(X\otimes_\Al\tilde\pi^o(\nabla^*\eta)|\psi\rangle\Bigr)&=&
-\tilde\pi(\nabla {\bf h}^\flat(\eta))\bigl(\J|\psi\rangle\Bigr)\otimes_\Al{\bf h}^\sharp(X)\;,\\
\J_{\bf h}\Bigl(\tilde\pi(\nabla X)|\psi\rangle\otimes_\Al \eta\Bigr)&=&
-{\bf h}^\flat(\eta)\otimes_\Al\tilde\pi^o({\bf h}^\sharp(\nabla X))|\psi\rangle\;,\eeann
it is seen that \(\J\Di=\epsilon^{\;\prime}\Di\J\) implies
\be\label{Cov14}
\J_{\bf h}\Di_{\nabla}=\epsilon^{\;\prime}\Di_{\nabla}\J_{\bf h}\;.\ee
Just as $\Hi$ is a Hilbert $\Al$-bimodule, $\Hi_{Cov}$ is a Hilbert $\End$-bimodule with the action of ${\bf A},{\bf B}\in\End$ given by :
\[X\otimes_\Al|\psi\rangle\otimes_\Al \eta\rightarrow({\bf A}X)\otimes_\Al|\psi\rangle\otimes_\Al (\eta{\bf B})\;.\]
The spectral data $\left\{\Al,\Hi,\Di,\chi,\J\right\}$, also called (non-commutative) geometry in \cite{Var}, is said to be Morita equivalent to the geometry defined by $\left\{\End,\Hi_{Cov},\Di_{\nabla},\chiral,\J_{\bf h}\right\}$.\\
Not only \(\Hi_{Cov}=\Mod\otimes_\Al\Hi\otimes_\Al\Mod^*\), but also \(\Hi=\Mod^*\otimes_\End\Hi_{Cov}\otimes_\End\Mod\) !
%%%%%%%%%%

\newpage
%%%%%%%%%%%%%%%%%%%%%%%%%

%%%%%%%%%%%%%%%%%%%%%%%%%
%%%%%%% TOT HIER IS HET GOED !!!!?????????
%%%%%%%%%%%%%%%%%%%%%%%%
%\poiju
\end{document}